# Higher order tensor decomposition for proportional myoelectric control based on muscle synergies

Ahmed Ebied [a,*], Eli Kinney-Lang [b], Javier Escudero [a]

[a] School of Engineering, Institute for Digital Communications, University of Edinburgh, Edinburgh EH9 3FB, United Kingdom
[b] University of Calgary, 2500 University Drive, NW Calgary, AB, Canada T2N 1N4

ABSTRACT

Muscle synergies have recently been utilised in myoelectric control systems. Thus far, all proposed synergy-based systems rely on matrix factorisation methods. However, this is limited in terms of task-dimensionality. Here, the potential application of higher-order tensor decomposition as a framework for proportional myoelectric control is demonstrated. A novel constrained Tucker decomposition (consTD) technique of synergy extraction is proposed for synergy-based myoelectric control model and compared with state-of-the-art matrix factorisation models. The extracted synergies were used to estimate control signals for the wrist's Degree of Freedom (DoF) through direct projection. The consTD model was able to estimate the control signals for each DoF by utilising all data in one 3rd-order tensor. This is contrast with matrix factorisation models where data are segmented for each DoF and then the synergies often have to be realigned. Moreover, the consTD method offers more information by providing additional shared synergies, unlike matrix factorisation methods. The extracted control signals were fed to a ridge regression to estimate the wrist's kinematics based on real glove data. The Coefficient of Determination ($R^2$) for the reconstructed wrist position showed that the proposed consTD was higher than matrix factorisation methods. In sum, this study provides the first proof of concept for the use of higher-order tensor decomposition in proportional myoelectric control and it highlights the potential of tensors to provide an objective and direct approach to identify synergies.

## 1. Introduction

Muscle synergy and the concept of modular organisation of muscle activity have been accepted as a framework to analyse the fundamental roles underlying the coordinated motor activity [1]. The muscle synergy concept would help to solve the complexity problem of motor control concerning the redundant number of actuators needed for motor activity [2,3]. The muscle synergy model suggests that the nervous system activates muscles in groups (synergies) for motor control rather than activating each muscle individually [4]. Muscle synergies has been proved to be an important analysis tool for many applications such as clinical research [5] and biomechanical studies [6,7] thanks to the fact that they can be extracted from non-invasive electromyography (EMG). According to the time-invariant synergy model [4,8], the estimation of muscle synergies and their weighting functions from a multi-channel EMG signal is a blind source separation (BSS) problem. Several matrix factorisation techniques have been used to solve this problem by estimating unknown synergies, with the non-negative matrix factorisation (NMF) algorithm [9] being the most prominent method [10,11]. However, in recent years, tensor decomposition has been introduced for synergy estimation [12], and a robust technique [13,14] has been developed for muscle synergy extraction. We now seek to explore the usefulness of this approach for proportional myoelectric control.

For decades, EMG has been used to control prostheses [15]. In addition to the conventional direct control approach, the current state-of-the-art methods for prosthetic upper-limb are usually based on pattern recognition techniques [16] which have been successful in achieving high classification accuracy for a range of motions (10 classes) [17]. Moreover, pattern recognition-based systems recently found their way into commercial products such as "Complete Control".[1]

However, pattern recognition systems generally provide sequential control schemes [18]. Natural limb movements consist in the simultaneous and proportional activation of multiple DoFs [19]. Thus, muscle synergies have been utilised in prosthesis control to achieve a simultaneous and proportional myoelectric control across multiple DoFs [20,21]. Most approaches for upper-limb synergy-based myoelectric control [22–24] rely on a matrix factorisation algorithm (usually NMF) to extract muscle synergies from a training multichannel EMG dataset. Then, the extracted synergies are used to estimate proportional and continuous control signals.

* Corresponding author. Present address: Biomedical Engineering Department, Military Technical College, Cairo, Egypt.
*E-mail addresses:* ahmed.m.ebied@gmail.com, ahmed.ebied@ed.ac.uk (A. Ebied).

[1] https://www.coaptengineering.com/.

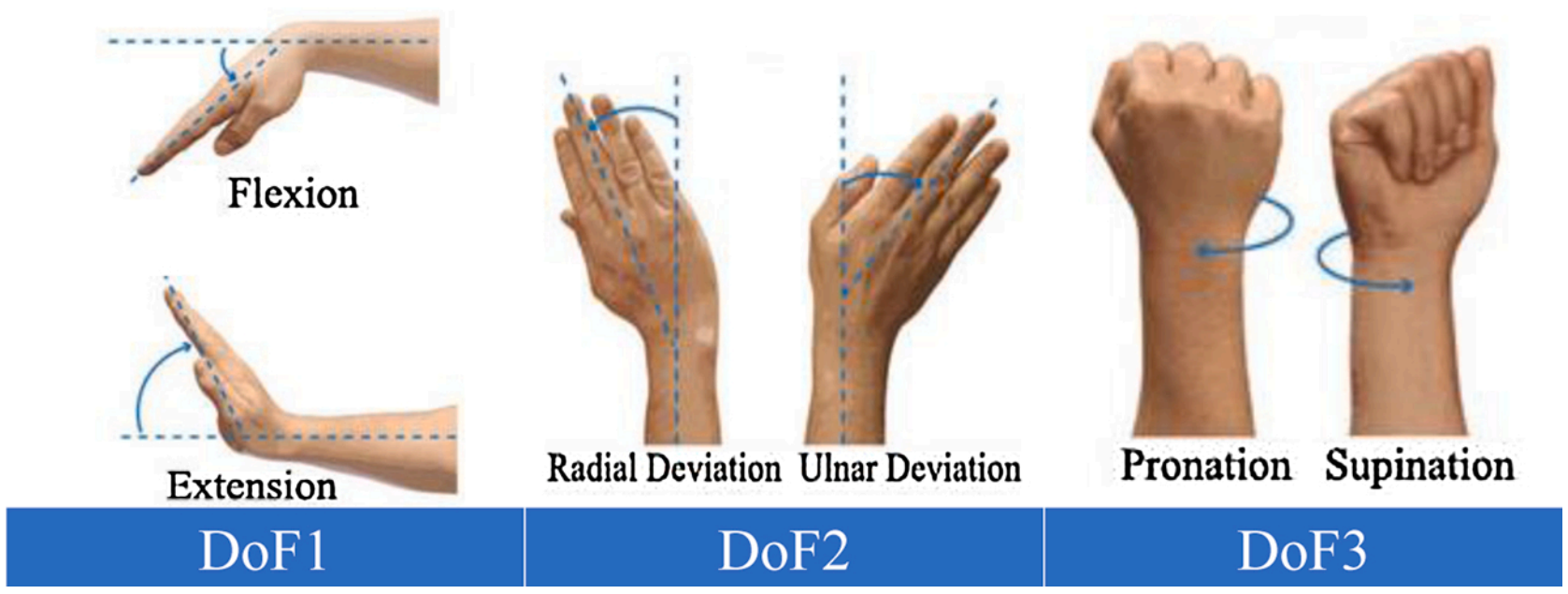

**Fig. 1.** The 6 movements selected to represent wrist's DoFs.

Synergy-based myoelectric control schemes need to identify the muscle synergies and their weighting functions associated with single-DoF. In this way, a control signal which corresponds to a Degree of Freedom (DoF) can be estimated through matrix factorisation. However, NMF is unable to extract the specified DoF synergies without further conditions imposed on the protocol. To tackle this problem, Choi and Kim [23] chose a completely supervised approach using a joint synergy matrix. Jiang et al. [20,22] proposed "divide and conquer" method, a semi-supervised approach which was used in [24] as well.

The SNMF approach is similar to the classical NMF method, but it tries to exploit the fact that some recent studies suggest the sparse nature of muscle synergies [11,25] and the lack of sparseness solution is one of the notable drawbacks for NMF [9,26]. Therefore, SNMF would help to improve the muscle synergy estimation and simplify the training stage as demonstrated by Lin et al. [21]. Recently, a similar approach using SNMF was introduced by Lin et al. [21]. This approach tries to exploit the fact that some recent studies suggest the sparse nature of muscle synergies [11,25], since the lack of sparseness solution is one of the notable drawbacks for NMF [9,26]. Therefore, SNMF would help to improve the muscle synergy estimation and simplify the training stage as demonstrated by Lin et al. [21]. SNMF was utilised to identify control signals from two DoFs training datasets where synergies are assigned to their respective DoF after matrix factorisation which makes it a quasi-supervised approach.

The performance of proportional myoelectric control based on NMF synergies degrades significantly with the increase in task-space dimension into three DoFs of movement [20,24]. In addition, the current approaches assign two synergies for each DoF (one synergy per movement). Thus, the number of synergies needed for control increases with the number of movements [27].

We hypothesise that tensor decomposition could help to solve this problem by incorporating the movement and DoF information into the decomposition process. Hence, the control signals for each DoF can be extracted directly with an appropriate tensor decomposition method. This is encouraged by our preliminary study [14] which showed that tensor decomposition was able to estimate consistent synergies when the task dimensionality is increased up to 3-DoFs, something that cannot be achieved via traditional matrix factorisation.

In a nutshell, higher-order tensors are the generalisation of matrices, which are 2nd-order tensors. Tensor decompositions provide several advantages over matrix factorisation such as compactness, uniqueness of decomposition, and generality of the identified components [28]. Moreover, EMG data are naturally structured in higher-order form in many applications, such as repetition of subjects and/or movements. Hence, matrix factorisation methods have limitations. For instance, in biomechanical studies, identifying shared muscle synergies requires to apply NMF repetitively to each movement and/or subject, then relying on metrics such as the correlation coefficient to identify the shared and task-specific synergies. This makes such an approach complex and unreliable [12]. Hence, tensor decomposition was utilised to identify muscle synergies through a variant of Tucker decomposition named constrained Tucker decomposition (consTD) [13]. Moreover, Takiyama et al. used parallel factor analysis (PARAFAC) decomposition for joint angle and EMG to estimate spatial, temporal and the task-specific synergies [29]. consTD was introduced as a framework for muscle synergy analysis [13] as it provided unique and consistent muscle synergies in comparison with unconstrained Tucker model. This proposed model was capable of identifying shared synergies across movements [14].

In this paper, the consTD method is proposed for proportional myoelectric control. The EMG data is tensorised by adding $\mathrm{movement}$ mode to the $\mathrm{spatial}$ ($\mathrm{Channels}$) and $\mathrm{temporal}$ ($\mathrm{time}$) modes to create a 3rd-order tensor with dimensions $\mathrm{time} \times \mathrm{channel} \times \mathrm{movements}$ (see Section 3.1.1 for details). Control signals are estimated from this tensor via consTD. To assess this approach, control signals are mapped to hand kinematics through ridge regression. The results will be compared with NMF and SNMF using two publicly available datasets. Therefore, this paper contributes a novel technique to use 3rd-order tensor decomposition in a synergy-based myoelectric control system. Tensor synergies have not been utilised in myoelectric control before.

## 2. Materials

Two datasets from the publicly available Ninapro [30,31] were used in this paper (http://ninapro.hevs.ch/node/7). The first dataset [32] consists of 27 able-bodied subjects instructed to perform 10 repetitions of 53 hand, wrist and finger movements. The dataset includes 10-channel EMG signals recorded by a MyoBock 13E200-50 system (Otto Bock HealthCare GmbH), rectified by root-mean-square (RMS) and sampled at 100 Hz. The hand kinematics were captured using a 22-sensor CyberGloveII (CyberGlove Systems LLC). The glove returns 8-bit values proportional to joint angles using a resistive bend-sensing technology with an average resolution of less than one degree depending on the size of the subject's hand. Data synchronisation was performed offline using high-resolution timestamps [30]. The "stimulus" time series in the Ninapro dataset labelled the start and end of each movement repeated by the subject. This series has been used for dataset segmentation of the training and testing datasets. The signals are divided into training and testing sets with 60% (six repetitions of each movement) of the data assigned to the training for each subject. The wrist motion and its three DoFs are investigated. Therefore, six movements are selected to represent the wrist's DoFs which are: the wrist radial and ulnar deviation that creates the horizontal DoF (DoF1); wrist extension and flexion movements which form the vertical DoF (DoF2); and finally wrist supination and pronation (DoF3) as shown in Fig. 1.

The second dataset [33] consists of 40 able-bodied subjects instructed to perform six repetitions of 50 hand, wrist and finger movements.

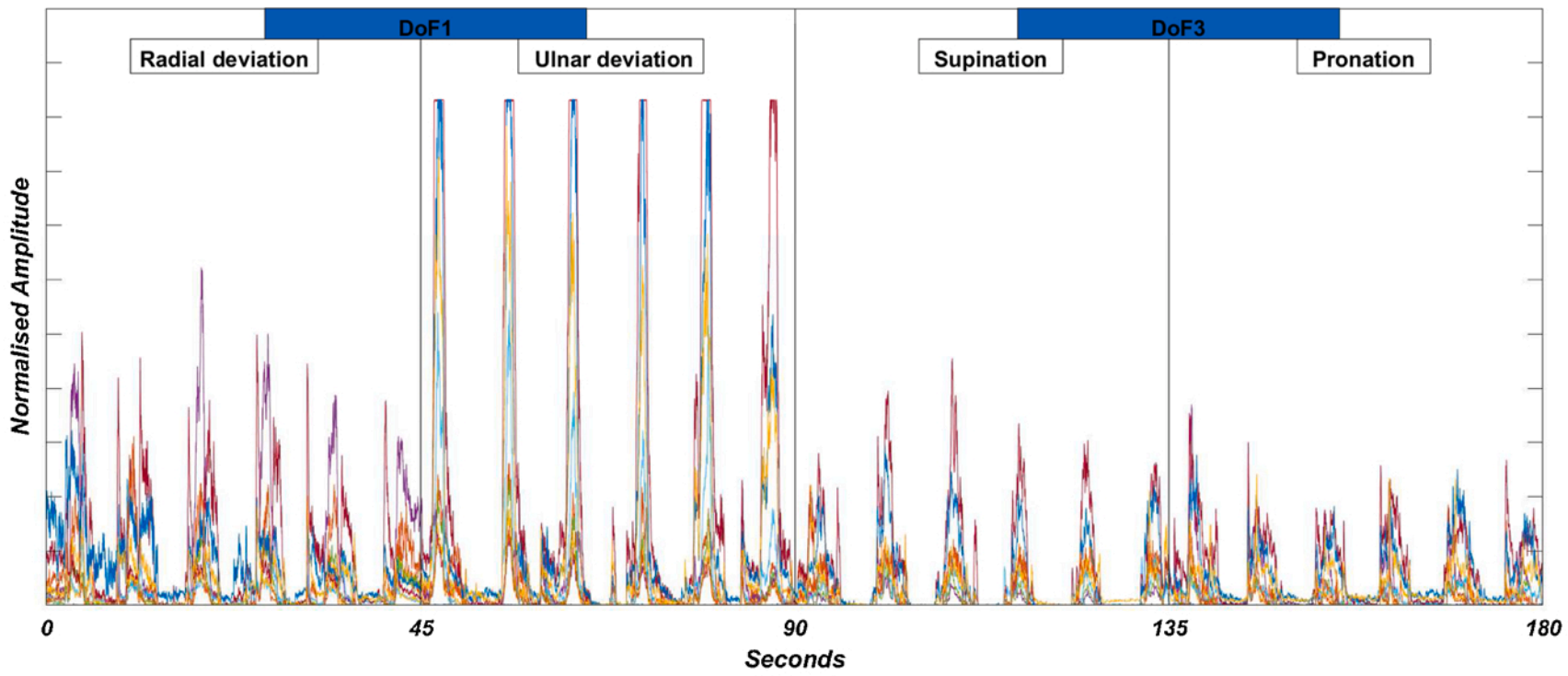

(a) *An example of the 10-channels surface EMG training dataset for DoF1-3. It consists of 6 repetitions for the 4 wrist's movements forming DOF1-3 (radial/ulnar deviation and supination/pronation).*

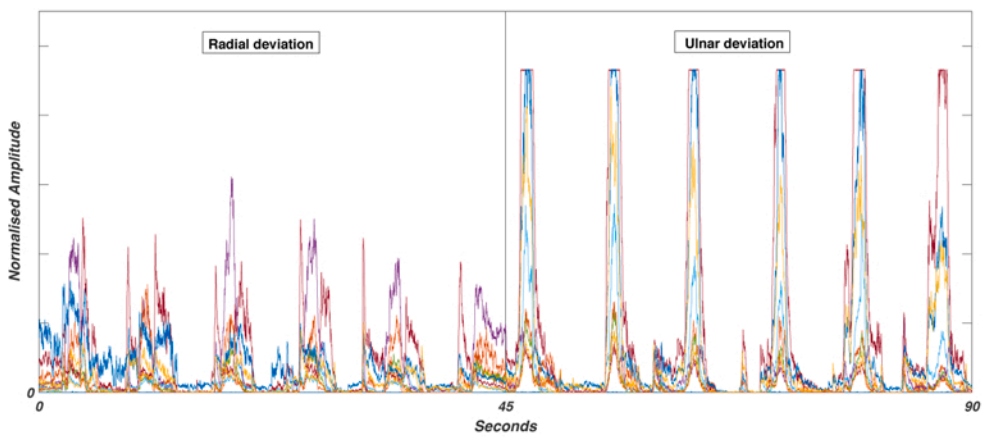

(b) *The data preparation for NMF and SNMF to estimate the muscle synergies for DoF1.*

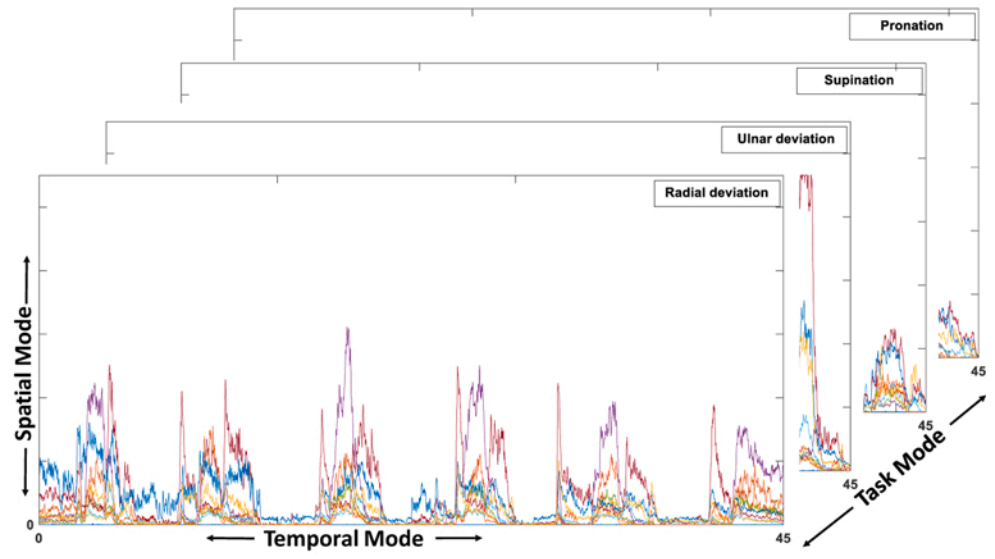

(c) *3$^{rd}$-order tensor for DoF1-3 with modes (time$\times$ Channels $\times$ movements).*

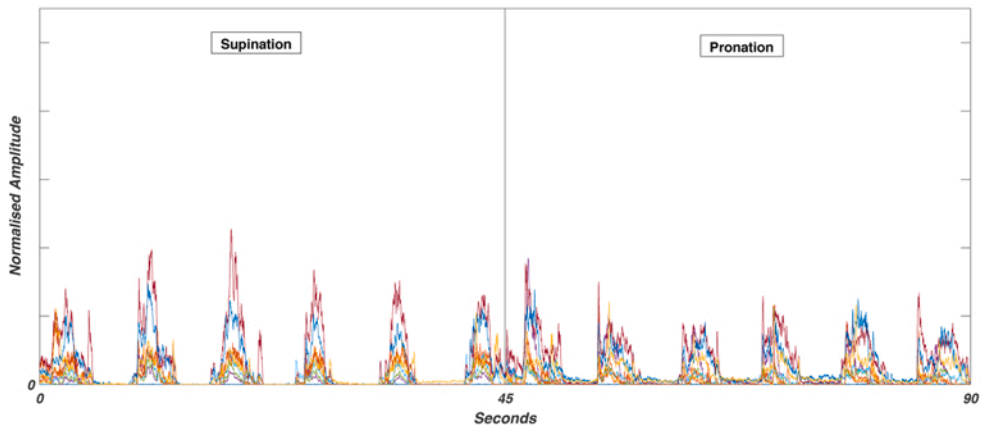

(d) *The data preparation for NMF and SNMF to estimate the muscle synergies for DoF1.*

**Fig. 2.** An example for training data preparation and tensor construction for subject 6 and DoFs 1 and 3. Panel a shows the whole recorded segment for the 6 training repetitions of the 4 movements. Data preparation for both NMF and SNMF methods are illustrated in Panels b and d, The data is divided into two separate segments for each DoF and NMF is applied to estimate 2 muscle synergies from each segment (1 for each movement). c shows the 3rd-order tensor construction by stacking the 4 movements in Panel a as separate slabs. Tensor decomposition is applied to directly estimate 6 synergies (4 task-specific and 2 shared).

The same wrist's movements investigated in the first dataset were selected from the second one. However, the myoelectric activity in this dataset is recorded with 12-channel setup by Delsys Trigno Wireless System. This different setup allows to record raw EMG signals sampled at 2 kHz with a baseline noise of less than 750 nV RMS. The EMG data is rectified by RMS in the pre-processing stage. Hand kinematics were captured using the same 22-sensor CyberGloveII system (CyberGlove Systems LLC) used in the first dataset. As mentioned, three wrist's DoFs are investigated with four repetitions of training and two assigned to the testing dataset.

## 3. Mathematical models

In this section, the mathematical models for synergy extraction approaches used in this paper will be described. We will present the higher-order tensor model (including the steps of constructing the tensor, the factorisation, and our novel method of consTD). Then matrix factorisation benchmarks will then be described. We will focus on NMF and SNMF as the current state-of-the-art synergy extraction methods. The difference between them as well as the difference between consTD will be highlighted.

### *3.1. Higher-order tensor models*

#### *3.1.1. Tensor construction*

The current muscle synergy extraction approaches prepare the EMG data in a matrix form with temporal and spatial dimensions. Hence, matrix factorisation methods are applied such as NMF and SNMF. Thus, in the case of synergy-based myoelectric control, where the EMG data consists of different movements and/or DoFs, synergies are extracted from each segment of movement or DoF separately [20,22].

For example, Fig. 2a shows a sample EMG data for six repetitions of four movements (two DoFs). In the case of matrix factorisation, the data is divided into two separate segments, one for each DoF as shown in Panels 2b and d. The synergistic information is estimated from each segment independently. On the other hand, the tensor decomposition prepare the data in a tensor form where all synergistic information are estimated from the same tensor.

3rd-order tensors are created by stacking the training EMG segments of each movement shown in Fig. 2a to form a tensor with modes; $\text{time} \times \text{channels} \times \text{movements}$ as shown in Fig. 2c. In this study, the training tensor is designed to have four different movements where a pair of them make a wrist's DoF. This results in three training tensors for each subject where each one consists of two wrist's DoF (four movements). The three tensors are named DoF1-2 for horizontal and vertical DoFs, named DoF1-3 for horizontal and inclination DoFs, and Finally, DoF2-3 for vertical and inclination DoFs.

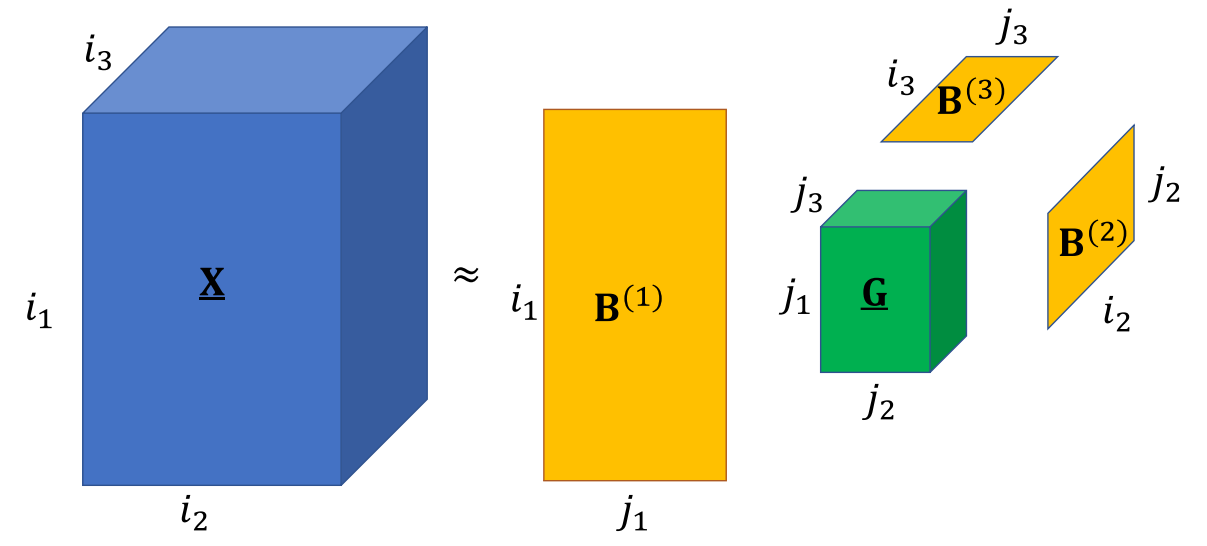

**Fig. 3.** Illustration of Tucker decomposition for 3rd-order tensor $\underline{X}$.

#### 3.1.2. Tucker decomposition model

Several decomposition models have been introduced to decompose higher-order tensors into their main components. Tucker decomposition [34] is one of the most prominent models for tensor factorisation [35]. In the Tucker model, an $n$th-order tensor $\underline{X} \in \mathbb{R}^{i_1 \times i_2 \times \cdots \times i_n}$ is decomposed into a smaller core tensor ($\underline{G} \in \mathbb{R}^{j_1 \times j_2 \times \cdots \times j_n}$) transformed by a matrix across each mode (dimension) [36], where the core tensor determines the interaction between those matrices as the following:

$$\underline{X} \approx \underline{G} \times_1 B^{(1)} \times_2 B^{(2)} \times \cdots \times_n B^{(n)} \tag{1}$$

where $B^{(n)} \in \mathbb{R}^{i_n \times j_n}$ are the components matrices transformed across each mode while "$\times_n$" is multiplication across the $n$th-mode [36]. The number of components for each mode ($j_n$) or the core tensor $\underline{G}$ dimensions is flexible (and they can be different) as long as ($j_n \leq i_n$). Tucker decomposition for a generic 3rd-order tensor is illustrated in Fig. 3.

The Tucker model usually uses the Alternating Least Squares (ALS) to estimate the core tensor and the component matrices. ALS has two main phases. The first one is the initialisation of components and core tensor either randomly or by certain criteria [37]. The second phase is a series of iterations to minimise the loss function between the original data and its model. For example, the least squares loss function for a 3rd-order Tucker model would be:

$$\underset{B^{(1)},B^{(2)},B^{(3)},\underline{G}}{\operatorname{argmin}} \| \underline{X} - B^{(1)}\underline{G}(B^{(3)} \otimes B^{(2)})^{T} \|^2 \tag{2}$$

where $\otimes$ is Khatri-Rao product which is the column-wise Kronecker product. This loss function is solved by fixing three out of four factors ($B^{(1)}, B^{(2)}, B^{(3)}$ and $\underline{G}$) and computing the unfixed factor by iterating alternatively. Although ALS has several advantages, its main drawback is that it cannot guarantee convergence to a stationary point [38] since the problem could have several local minima. This can be solved by applying multiple constraints on the initialisation and iteration phases [39] to improve the estimation. Hence, constraints over both initialisation and iteration phases can help to solve the convergence problem. Moreover, the constrained Tucker model has several benefits including: uniqueness of the solution, and interpretable results that do not contradict prior knowledge and finally speeding up the algorithm. Although constraints could lead to poorer fit of the data compared to the unconstrained model, the advantages outweigh the decrease in the fit for most cases [28].

Therefore, in our previous work [13], we developed a constrained Tucker model (consTD) that was able to estimate a unique and interpretable shared and task-specific synergies with high explained variance and short execution time in comparison with the standard NMF approach.

#### 3.1.3. Constrained Tucker decomposition

Here, we devise the consTD for the extraction of muscle synergies that could be utilised in myoelectric control. This builds on top of our previous work on tensor models to extract muscle synergies [13,14] but, crucially, we now use 3rd-order EMG tensor data where the third additional mode is movements instead of repetitions, as described in Section 3.1.1. This change in the tensor construction was implemented so that consTD is applied to a data structure similar to the matrix factorisation approaches under comparison. NMF or SNMF are applied on EMG segments of several repetitions and not on each repetition separately, something that will be discussed in detail in Section 3.2. Moreover, we test the ability of the proposed consTD to work with different settings and data construction.

Two constraints are imposed on the initialisation phase and one constraint in the iteration phase. For initialisation, the core tensor is initialised and fixed at a value of 1 between each component in the (temporal\movements) modes and its respective spatial synergy and 0 otherwise as the following:

$$\begin{aligned}
g_{n,n,n} &= 1 \quad n \in \{1,2,3,4\}, \\
g_{n,5,n} &= 1 \quad n \in \{1,2\}, \\
g_{n,6,n} &= 1 \quad n \in \{3,4\}, \\
g_{i,j,k} &= 0 \quad \text{otherwise.}
\end{aligned}$$

This core set-up that does not update with every iteration avoids undesired cross interactions between spatial components (synergies) and other modes components. The values in the core tensor are chosen to be 1 in order to hold any variability in the components rather than core tensor.

The second initialisation constraint fixes the movement mode components since we have the information about each factor and its corresponding movement. The values are designed to be 1 for the considered movement and 0 otherwise. Non-negativity constraint on temporal and spatial modes is the only constraint in the iteration phase. This is imposed to have meaningful factors (synergies) [23,12]. Non-negativity is a common constraint because of the illogical meaning of negative components in many situations. Here, it is beneficial due to the additive nature of muscle synergies. It is implemented in the iteration phase by setting the negative values of computed components to zero by the end of each iteration to force the algorithm to converge into a non-negative solution. A similar constrained set-up have been used in a previous study [13] to extract shared muscle synergies. Moreover, the algorithm runs for a minimum of ten iterations to ensure that the model does not converge to a poor local minimum and the decomposition with the highest explained variance is chosen.

This consTD approach would result in four task-specific synergies and two additional shared DoF synergies in the spatial mode. The additional DoF synergies are a shared synergy between the two movements (tasks) that form DoF. This is determined by the set-up of the core tensor for the 5th and 6th factors (synergies) as shown in Fig. 7.

### 3.2. Matrix factorisation models

To evaluate the tensor-based approach for proportional myoelectric control, we introduce NMF and SNMF as state-of-the-art benchmarks to compare with. In general, matrix factorisation is the standard approach for synergy extraction with NMF being the most prominent technique [11]. Both matrix factorisation methods – NMF and SNMF – are the main extraction methods used for synergy based myoelectric control [21].

#### 3.2.1. NMF

NMF [9] has been the most prominent method to extract muscle synergies [11]. In addition, it has been utilised for a proportional myoelectric control approach based on muscle synergies [20]. NMF processes the multichannel EMG recording as a matrix $X \in \mathbb{R}^{m \times n}$ with modes (channel $\times$ time). This matrix is factorised into two smaller

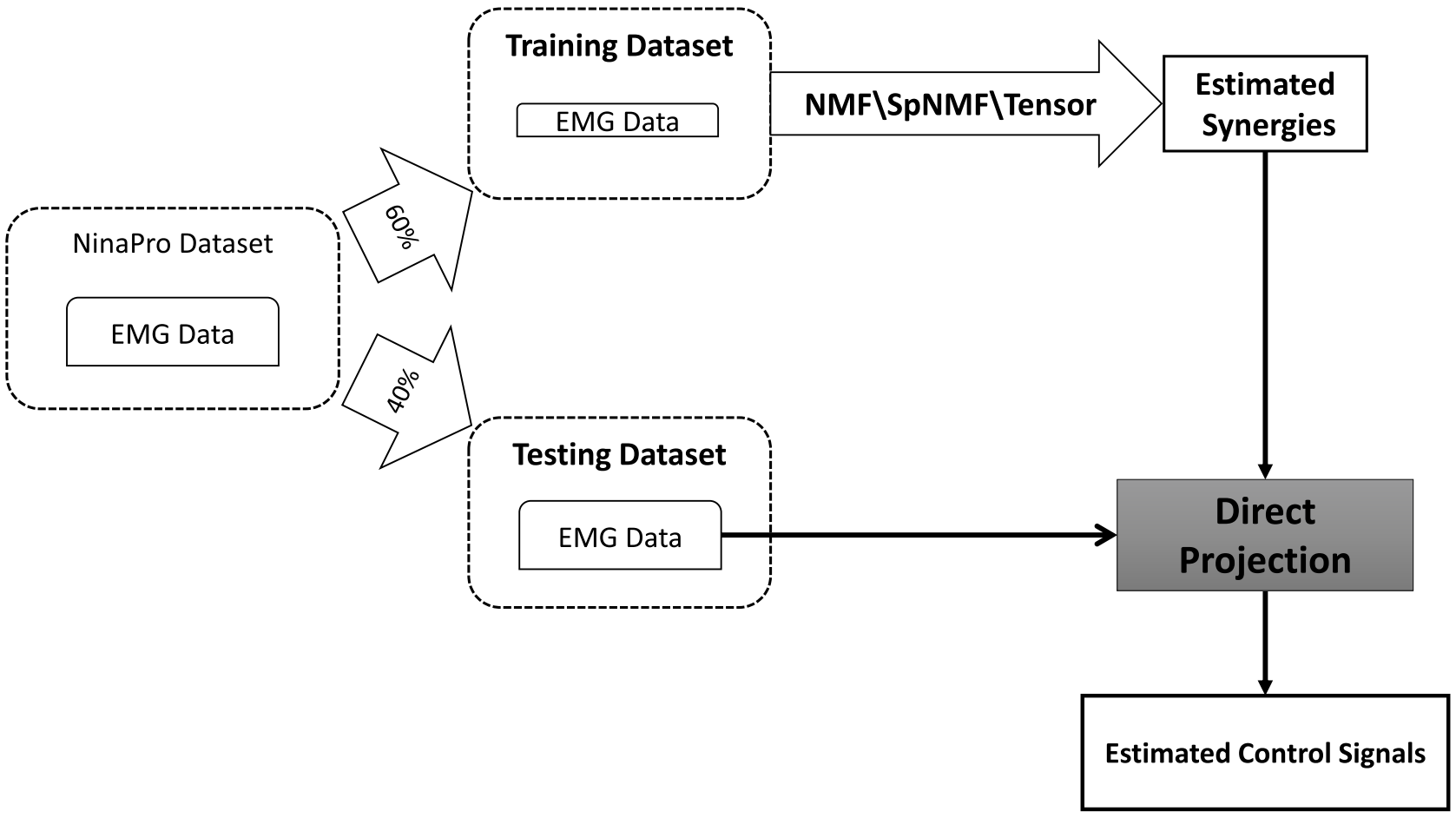

**Fig. 4.** Block diagram for the use of extracted muscle synergies from the training dataset to estimate control signals.

matrices (factors) as

$$X_{(m\times n)} = B^{(1)}_{(m\times r)} \times B^{(2)^T}_{(n\times r)} \tag{3}$$

where $B^{(1)} \in \mathbb{R}^{m\times r}$ holds the temporal information (known as weighting function) while the other factor $B^{(2)} \in \mathbb{R}^{r\times n}$ is the muscle synergy holding the spatial information and $r$ is number of synergies where $r < m, n$ to achieve dimension reduction. The algorithm relies on a cost function where both factors are updated and optimised with respect to the non-negativity constraint to minimise the difference between the data matrix X and its approximation as the following:

$$\begin{aligned} &\min_{B^{(1)},B^{(2)}} \frac{1}{2} \| X - B^{(1)}B^{(2)} \|_F^2 \\ &s.t. \quad B^{(1)}, B^{(2)} \geq 0 \end{aligned} \tag{4}$$

where $\| .\|_F$ is the Frobenius norm and both factors $B^{(1)}$ and $B^{(2)}$ are constrained to be non-negative. For more details, see [40].

In order to use the NMF synergies for a simultaneous and proportional myoelectric control scheme, Jiang et al. [20,22] proposed a "divide and conquer" approach. This is done by designing an experimental protocol to estimate muscle synergies and their respective weighting functions for a single DoF (two movements) at a time. Consequently, this approach would limit the factorisation into a few possible solutions. The result would be two muscle synergies and their respective weighting functions (or control signal) for each DoF, which allows simultaneous and proportional EMG control without multi-DOF training data.

#### 3.2.2. SNMF

The SNMF approach is similar to the classic NMF method, but it tries to impose sparseness constraints on the factorisation since the lack of sparseness solution is one of the notable drawbacks for NMF [9,26]. This is done by imposing a sparseness constraint to the weighting functions (control signals) based on the SNMF scheme introduced in [41]. In the case of SNMF algorithm, the cost function of classic NMF is shown in Eq. (4) is modified to the following:

$$\begin{aligned} &\min_{B^{(1)},B^{(2)}} \frac{1}{2} \| X - B^{(1)}B^{(2)} \|_F^2 + \lambda \sum_{n}^{j=1} \| B^{(2)}(:,j) \|_1^2 \\ &s.t. \quad B^{(1)}, B^{(2)} \geq 0 \end{aligned} \tag{5}$$

where $B^{(2)}(:,j)$ is the $j$th column vector of $B^{(2)}$ and $\lambda > 0$ is a regularisation parameter to balance the trade-off between the accuracy of the approximation and the sparseness of $B^{(2)}$ (control signals).

## 4. Methods

In this section, the methodology of comparing and assessing the use of consTD and matrix factorisation methods for synergy-based myoelectric control is discussed. All decomposition and computing are performed using **Matlab** 9 with *Intel* core i7 processor (2.4 GHz, 12 GB RAM). The consTD algorithm uses the "*tucker*" function in the N-way toolbox [42].

### 4.1. Muscle synergy extraction

To assess and compare between the application of tensor decomposition and matrix factorisation in a synergy-based myoelectric control system, muscle synergies were extracted from the EMG dataset.

The $\{4, 6, 4\}$ consTD method discussed in 3.1.3 decomposes the 3rd-order tensors to estimate the muscle synergies. The decomposed tensor consists of a pair of wrist's DoFs. For example, tensor (DoF1-3) of subject six is shown in Panel 2c. The tensor is decomposed into $\{4, 6, 4\}$ components across its three modes (temporal, spatial and movements) respectively. The first four components in the spatial mode are task-specific synergies for each movement of DoFs 1 and 3, while the 5th and 6th synergies are shared between DoFs 1 and 3 respectively. Those synergies are then used to estimate control signals through direct projection.

On the other hand, matrix factorisation methods, NMF and SNMF, decompose EMG segments of one DoF (two movements) into two synergies and their respective weighting functions. An example of EMG segments for subject six (DoFs 1 and 3) are shown in Panels 2b and d. This was applied to the three main wrist's DoFs separately. Then the extracted synergies are used to estimate the control signal through direct projection.

### 4.2. Direct projection of control signal

The identified synergies either from matrix factorisation methods or consTD are used to estimate the control signals as shown in Fig. 4.

The muscle synergies extracted using consTD on the training tensors are utilised to estimate one control signal per movement (four in total). This is done through direct projection of the testing data onto the fixed training components (core tensor and spatial\movement modes) to estimate the temporal mode components of the testing dataset. For the 3rd-order tensor in this study, the projection of the training DoF tensor $\underline{X}$ to the time mode ($B^{(1)}$) based on Eq. (1) would be

$$B^{(1)} = \underline{X}^{(i_1\times i_2 i_3)}[\underline{G}^{(j_1\times j_3 j_2)}(B^{(3)} \otimes B^{(2)})^T]^+ \tag{6}$$

where $B^{(2)}$ and $B^{(3)}$ are the spatial (synergy) and movements modes,

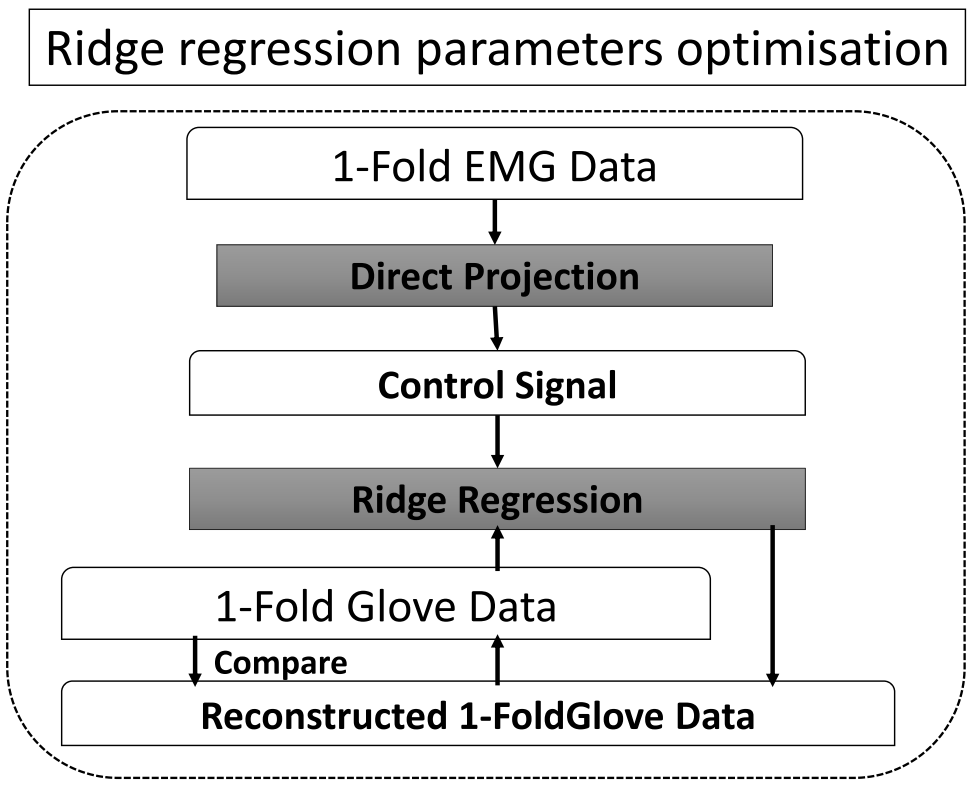

**Fig. 5.** The 10-fold cross validation process to optimise Ridge regression parameters.

respectively. Both modes are calculated from the training dataset, while $\underline{G}^{(j_1\times j_3 j_2)}$ is the fixed core tensor unfolded across the temporal mode ($j_1$). Therefore, Eq. (6) can be used to project the testing dataset ($\underline{X}_{test}$) to estimate the control signals (temporal mode) projection $B^{(1)}_{test}$. The resulting projection is the *time* mode matrix $B^{(1)}$. This projected matrix consists of four control signals, where each one represents the projection of one movement of the input testing dataset. The final control signal will be the difference between the two control signals of the two movements that form each DoF. Thus it could be used in real-time for myoelectric control.

In the case of matrix factorisation methods – either NMF or SNMF – control signals for each movement are estimated using the inverse model of the weighting functions. According to Eq. (3), the control signal C would be

$$C = X_{test} \times B^{(1)^{+}} \tag{7}$$

where $B^{(1)^{+}}$ is the pseudoinverse of the synergy matrix $B^{(1)^{T}}$ and $X_{test}$ is the testing EMG dataset for one DoF. The resulting projection consists of two control signals representing the projection of both movements of the DoF test dataset.

For myoelectric control applications, the final control signal is calculated for each DoF in a similar approach to other synergy-based myoelectric control studies [20,21]. It is deduced by taking the difference between the control signals of each movement and its antagonistic movement for each DoF. As a result, we estimate the final control signal for each wrist's DoF using NMF, SNMF and consTD methods.

### 4.3. Mapping into glove data

To demonstrate how our approach can be used in simultaneous and proportional myoelectric control systems, the testing EMG dataset is used to reconstruct its respective glove data using the estimated control signals from NMF, SNMF and consTD. The signals are mapped via ridge regression into the 22 sensor glove dataset as shown in Fig. 6. For all subjects, the reconstructed glove data is compared to the true testing dataset, where Coefficient of Determination ($R^2$) is calculated as an index for the quality of reconstruction.

The four control signals are regressed onto the 22 glove sensors data [43]. The coefficients for the multi-linear ridge regression are estimated separately from the training dataset of the same subject, then applied to the control signal to predict each glove sensor signal. The multi-linear ridge regression model estimates regression coefficients $\widehat{\beta}$ using

$$\widehat{\beta} = (X^T X + kI)^{-1} X^T y \tag{8}$$

where X is the predictor matrix and $y$ is the observed response. The regression parameter $k$ is a regularisation constant. To optimise these parameters, a 10-fold cross-validation (CV) procedure is designed. The training dataset for each subject is divided into 10 folds. For each fold, the optimisation of $k$ parameter is performed via a log-linear search to maximise the quality of regression using the $R^2$ index. The glove data were reconstructed using the muscle synergies and control signals estimated from the training datasets using the three methods under investigation as shown in Fig. 5. The $k$ regularisation constant parameter and regression coefficients $\widehat{\beta}$ were calculated from the training datasets and used to map the control signals of the testing data sets into the glove data to be compared.

To rule out any statistical chance from the comparison, random synergies are used to project control signals and regress the glove data as the other three methods. For each DoF, two random synergies are created from random values selected from a uniform distribution be-

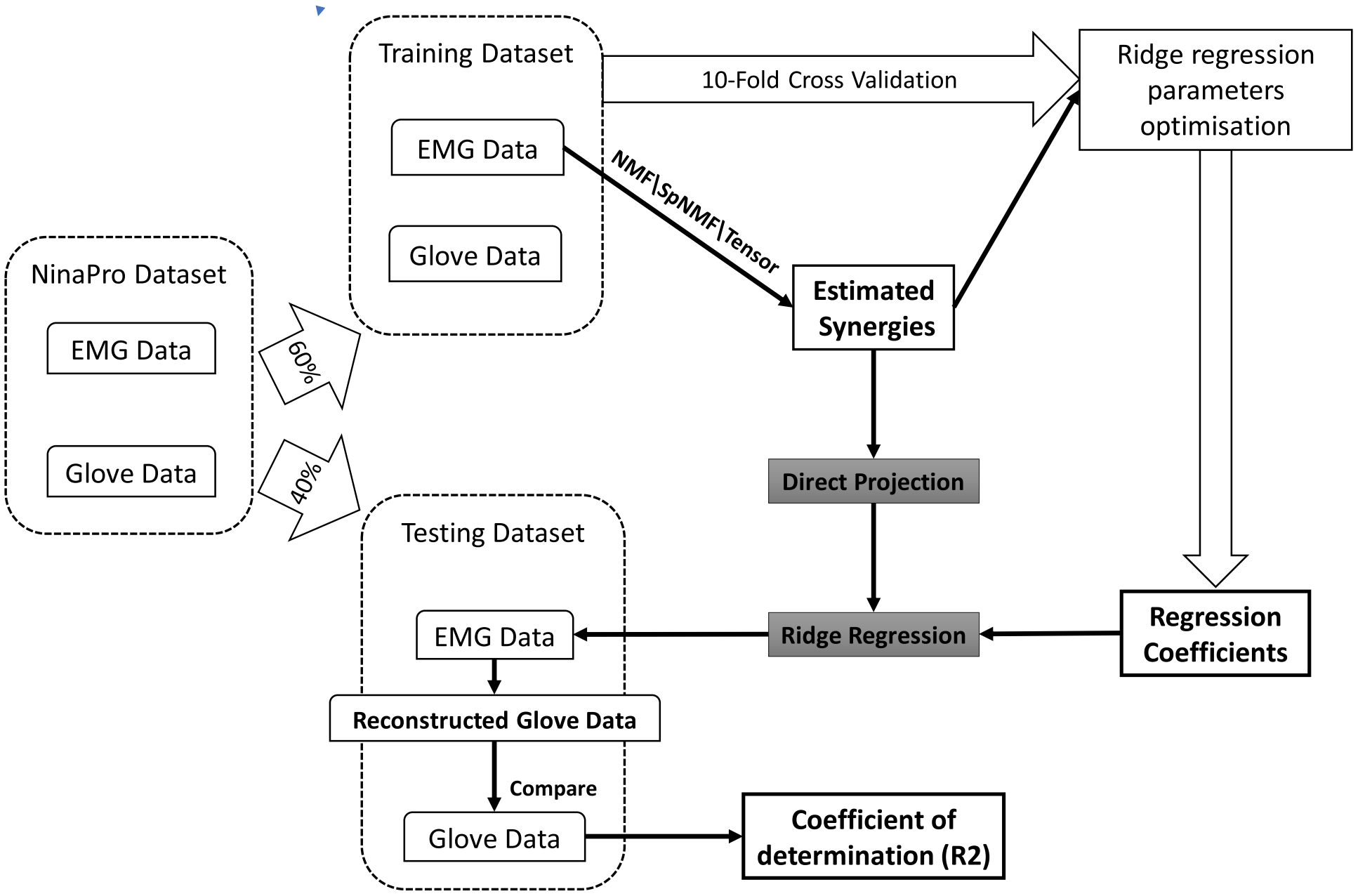

**Fig. 6.** Block diagram for the use of estimated synergies from the training dataset in reconstructing the glove testing dataset.

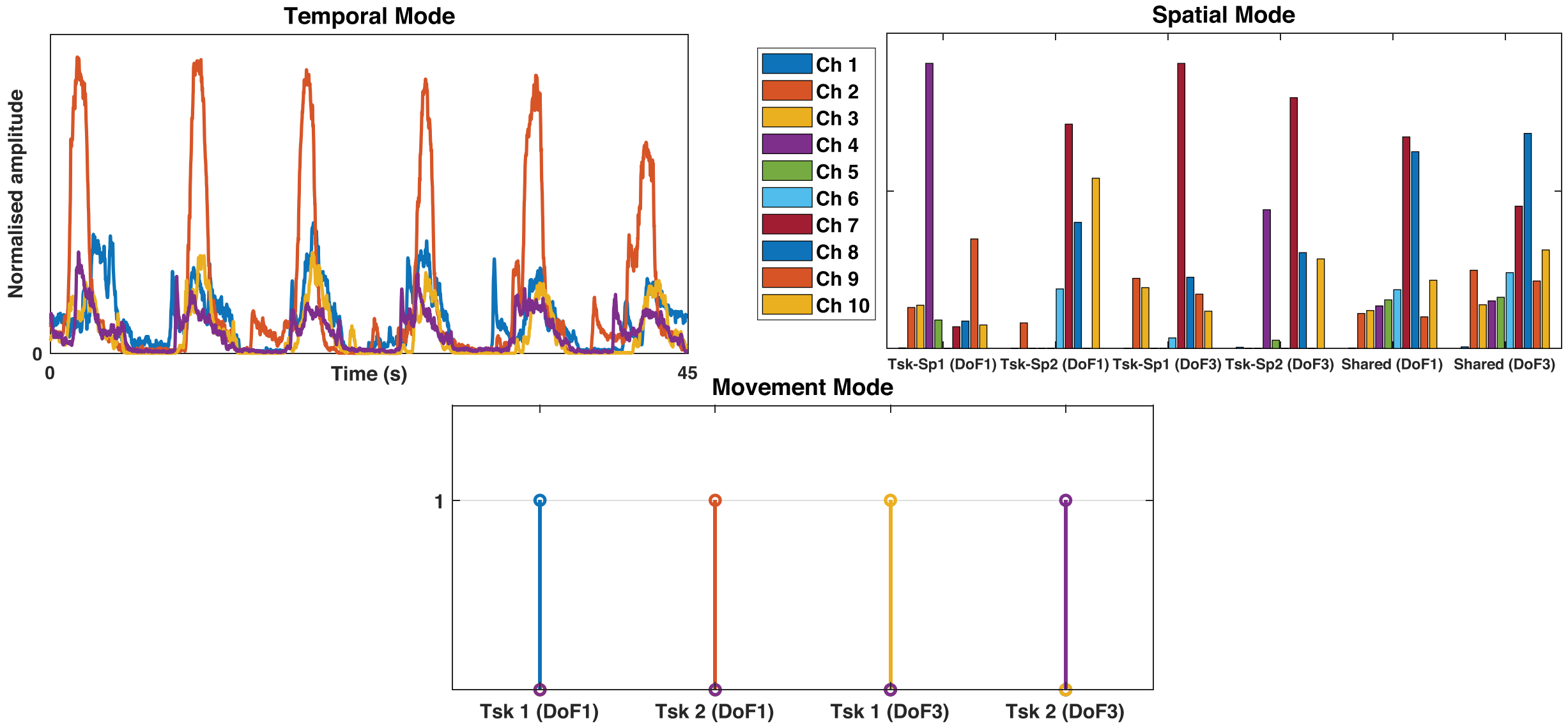

**Fig. 7.** consTD applied on DoF1-3 Tensor (shown in Fig. 2c). Spatial synergies 1 and 2 are specific for DoF1 while synergies 3 and 4 are specific for DoF3. Synergies 5 and 6 are shared synergies between DoFs 1 and 3.

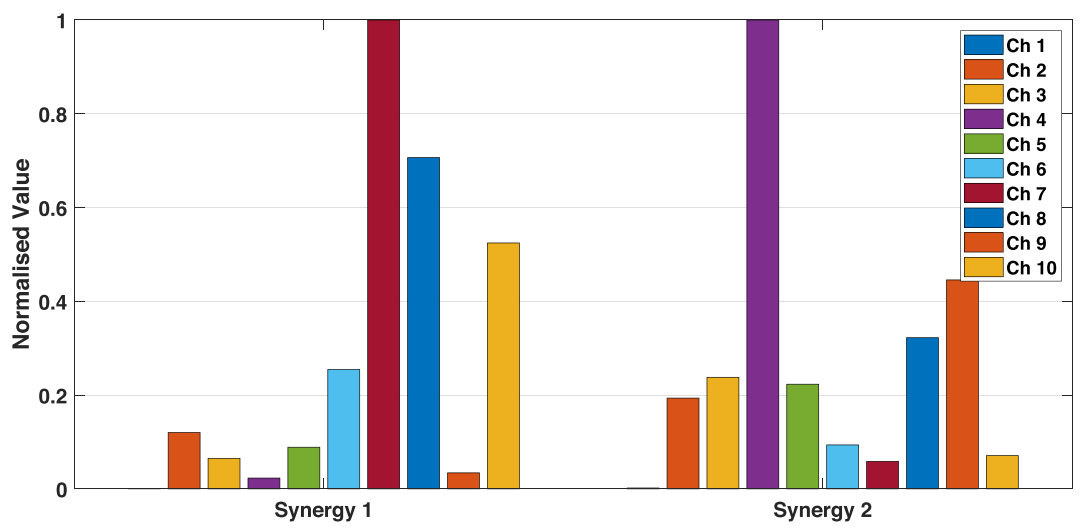

(a) *NMF synergies for DoF1.*

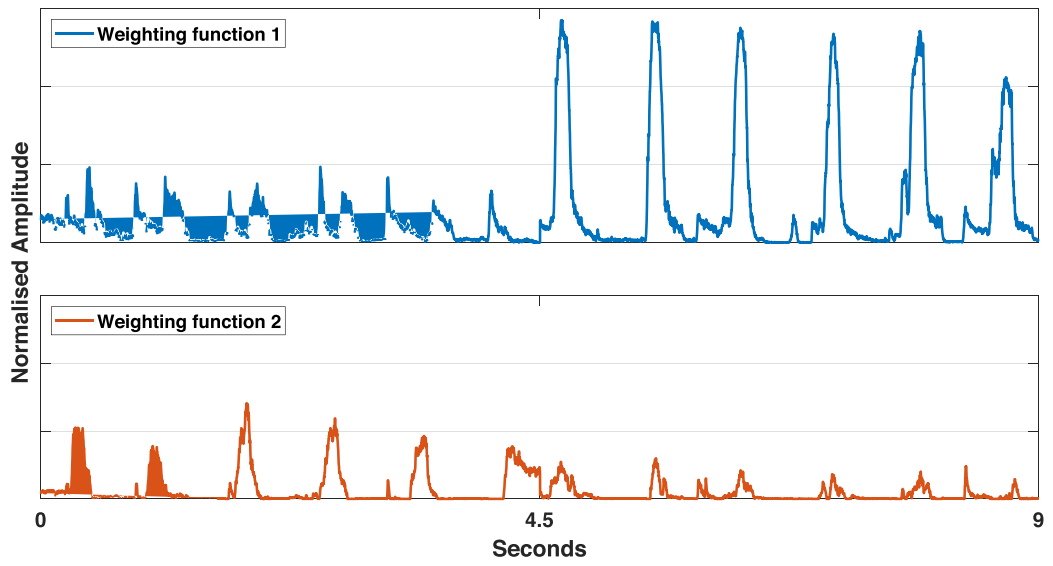

(b) *Weighting Functions for DoF1.*

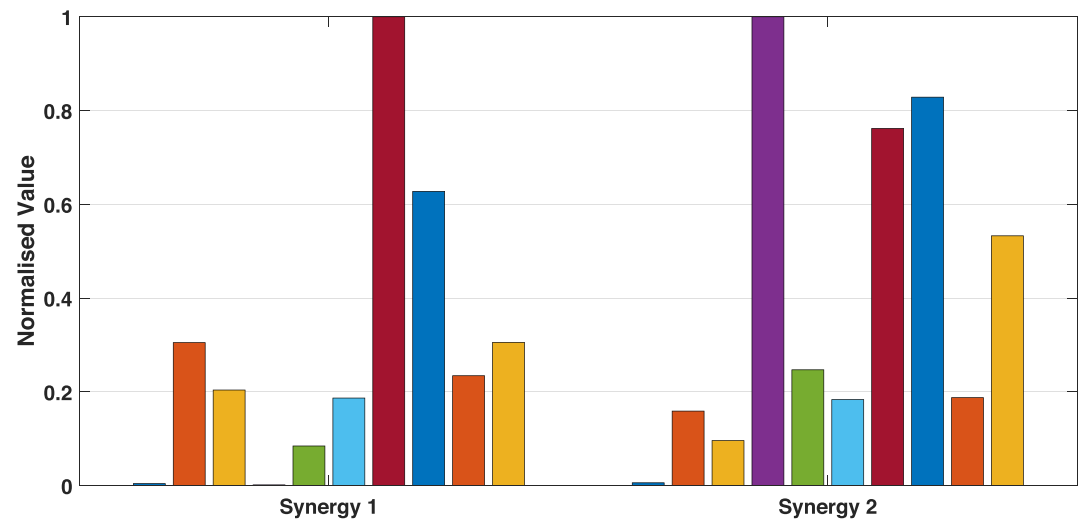

(c) *NMF synergies for DoF3.*

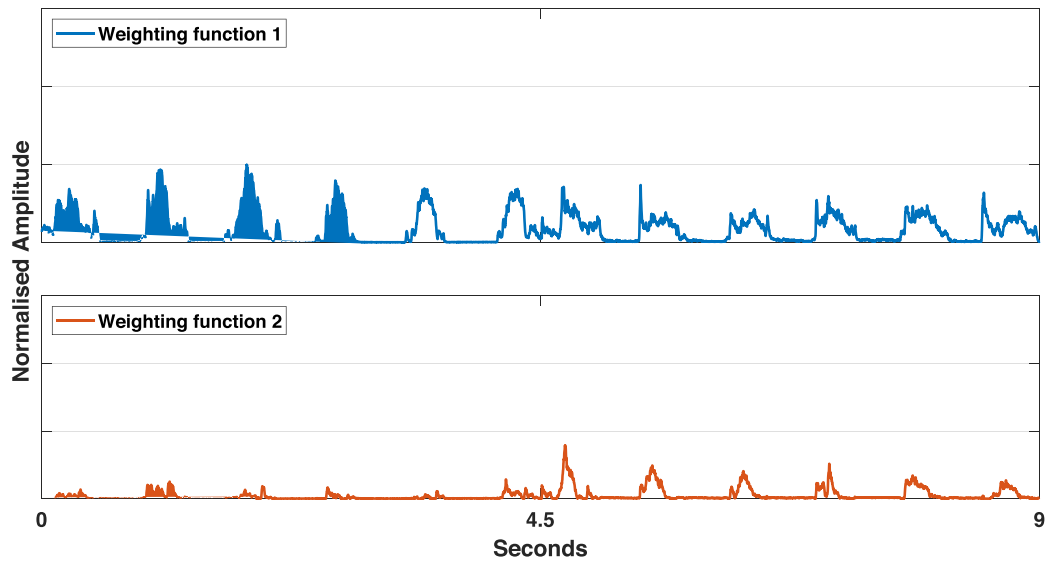

(d) *Weighting Functions for DoF3.*

**Fig. 8.** The NMF of training EMG datasets for DoF1 (Panels a, b) and DoF3 (Panels c, d) recorded from subject 6.

tween [0,1]. Two-sample *t*-test were conducted to compare the total $R^2$ of each technique and the randomly generated synergies.

Finally, since many of the 22 glove sensors are redundant and most of them do not capture the wrist's motion, the top three sensors across all methods for $R^2$ values are selected to represent the hand kinematics and were compared across all subjects.

## 5. Results

### 5.1. Constrained Tucker decomposition

The consTD decomposes the 3rd-order tensors constructed for each pair of wrist's DoFs. An example of the consTD for the EMG tensor (DoF1-3) of subject six is shown in Fig. 7. The tensor is decomposed into $\{4,6,4\}$ components across its three modes (temporal, spatial and movements) respectively where the core tensor and movement mode are constrained as discussed in detail in 3.1.3. Each component in the temporal mode is related to one movement of the four movements of DoFs 1 and 3. For the spatial mode, the first four components are task-specific synergies for those four movements, while the 5th and 6th synergies are shared synergies between wrist's DoFs 1 and 3 respectively.

Those task-specific and shared synergies are then used to estimate the control signals for the testing dataset through direct projection, as discussed in Section 4.2. An example of the final control signals for DoF1 and DoF3 of subject six estimated using the consTD approach are illustrated in Fig. 12.

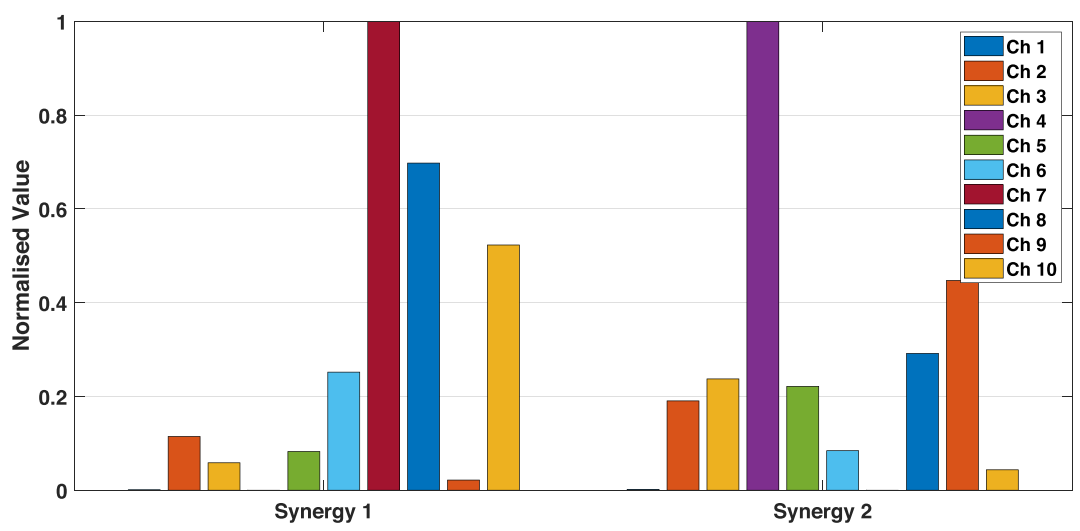

(a) SNMF synergies for DoF1.

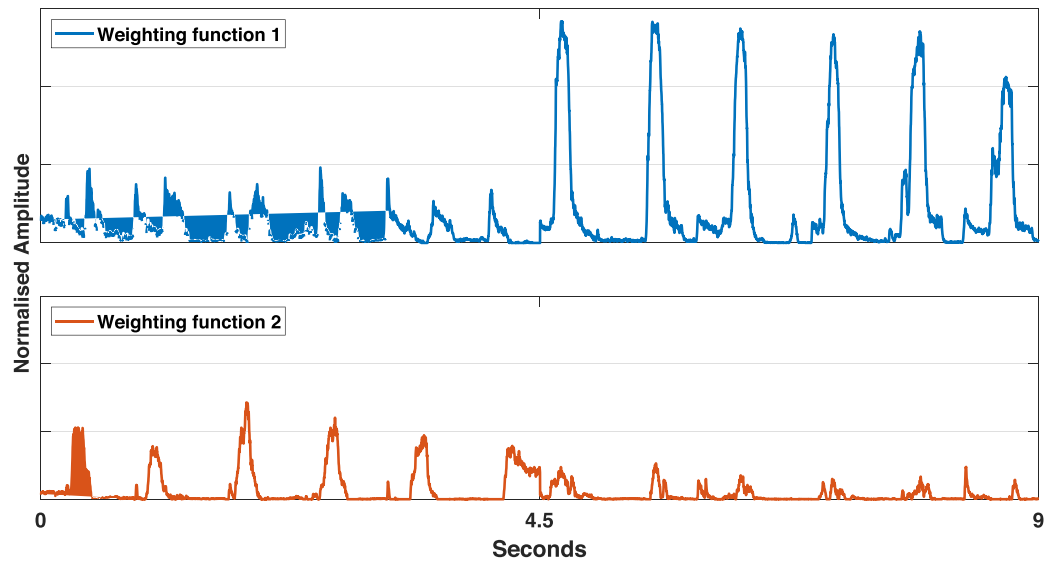

(b) *Weighting Functions for DoF1.*

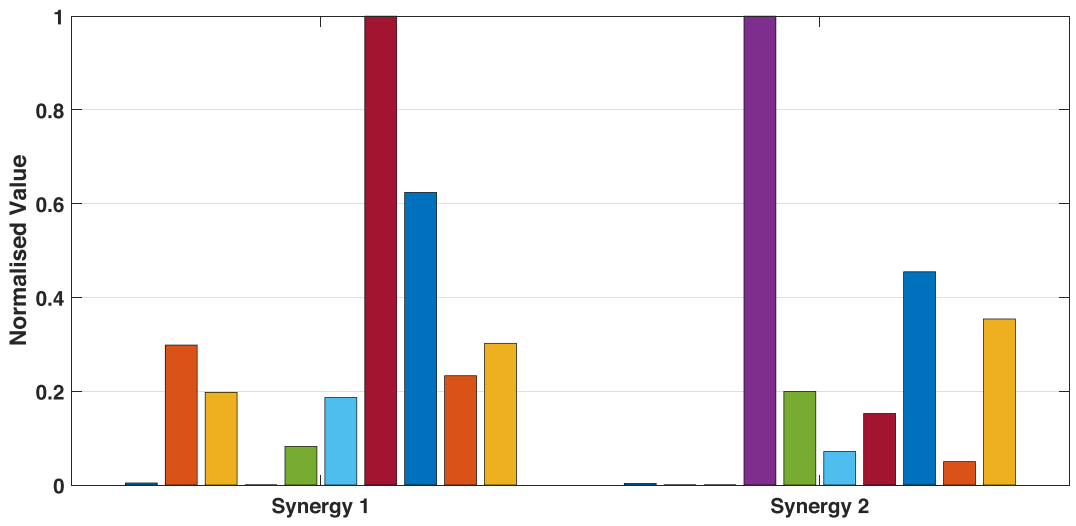

(c) *SNMF synergies for DoF3.*

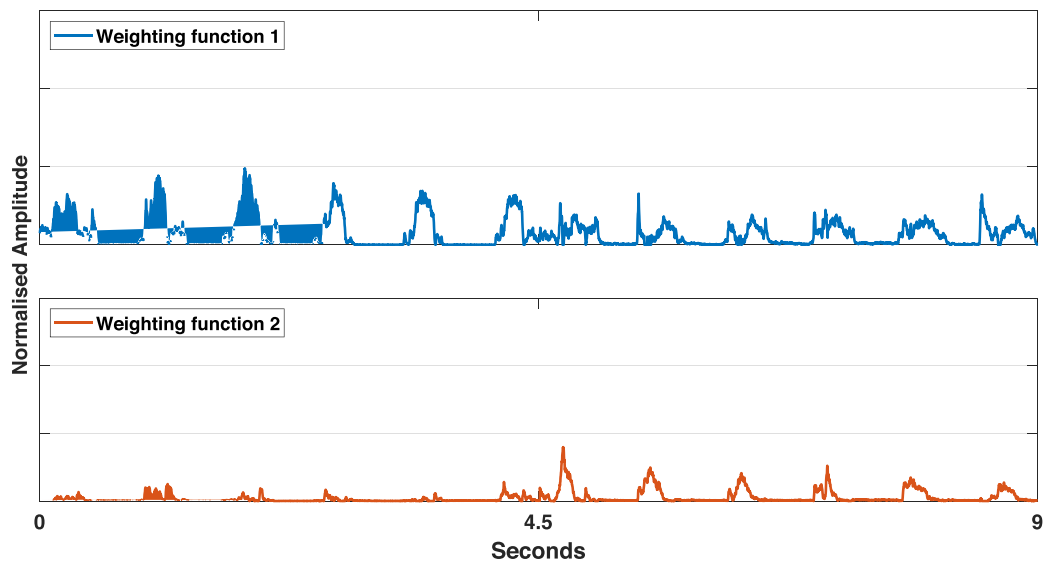

(d) *Weighting Functions for DoF3.*

**Fig. 9.** The NMF of training EMG datasets for DoF1 (Panels a, b) and DoF3 (Panels c, d) recorded from subject 6.

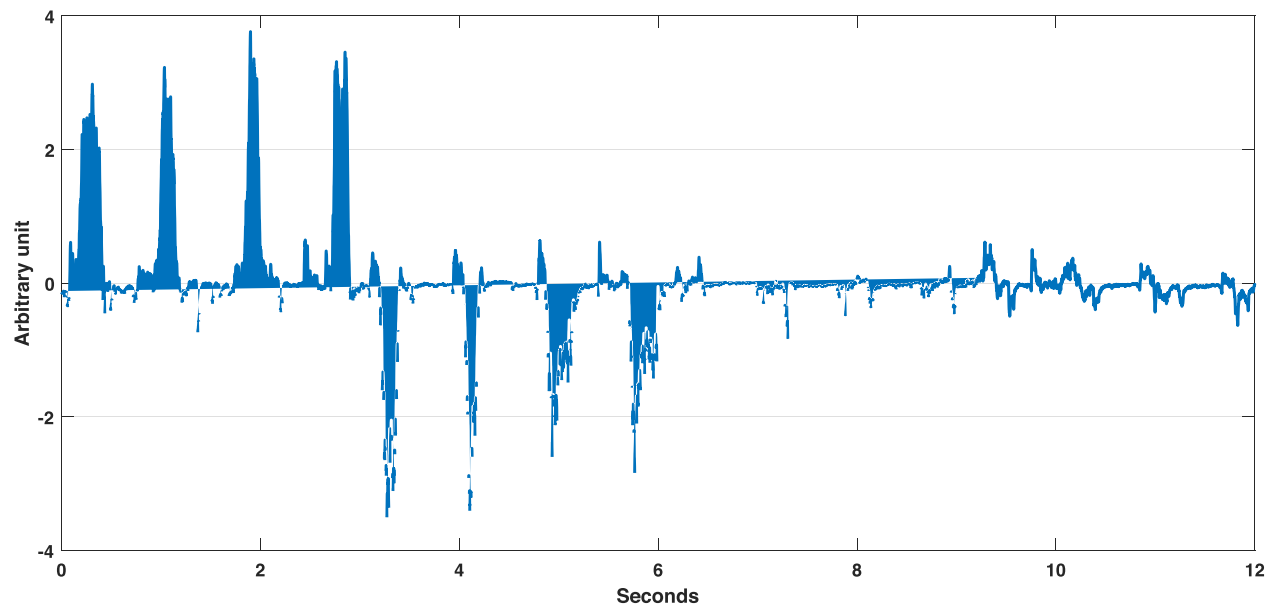

(a) *Final control signal for DoF1 using NMF synergies.*

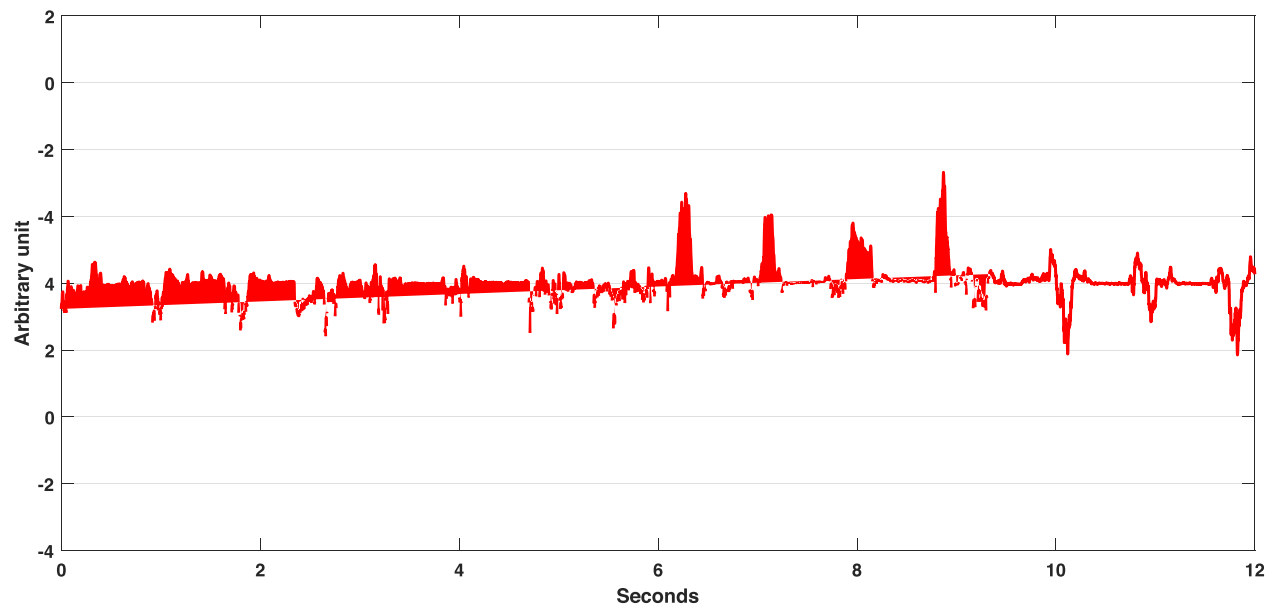

(b) *Final control signal for DoF3 using NMF synergies.*

**Fig. 10.** Final control signal for DoF1 (Panel a) and DoF3 (b) projected through direct projection of muscle synergies extracted via NMF recorded from subject 6.

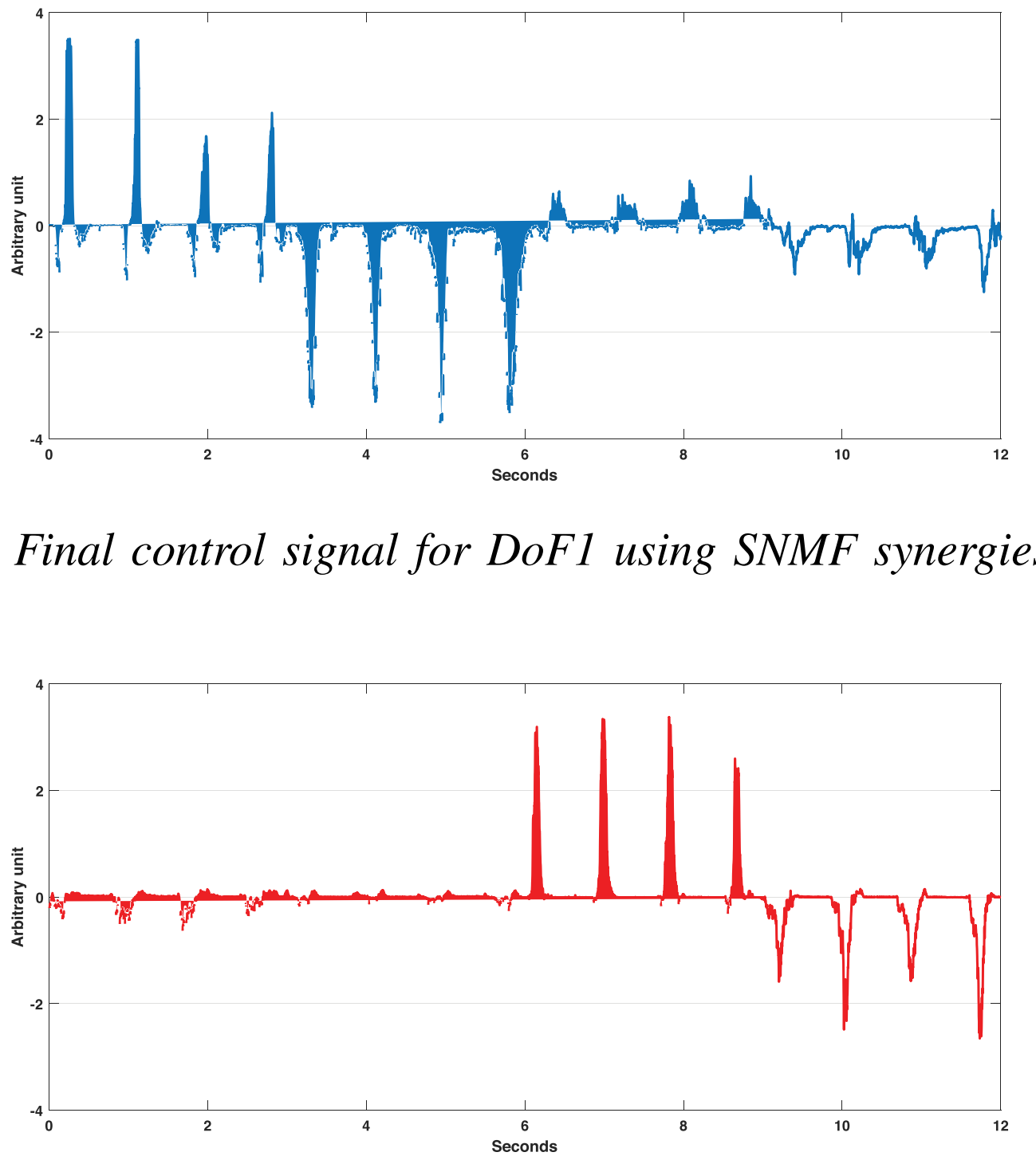

(a) *Final control signal for DoF1 using SNMF synergies.*

(b) *Final control signal for DoF3 using SNMF synergies.*

**Fig. 11.** Final control signal for DoF1 (Panel a) and DoF3 (b) projected through direct projection of muscle synergies extracted via SNMF recorded from subject 6.

### *5.2. Matrix factorisation models*

Both NMF and SNMF decompose a training EMG segment of one DoF (two movements) into two synergies and their respective weighting functions. This was applied to the three main wrist's DoFs separately. Then the extracted synergies were used for estimating the testing glove dataset through direct projection of EMG dataset. The SNMF was used to separate between movements directly by imposing sparseness on the weighting function. An example of NMF of DoF1 and DoF3 for subject six is shown in Fig. 8. The same segments were decomposed by SNMF as illustrated in Fig. 9.

Control signals for each movement are estimated using direct projection of the matrix factorisation components and consTD methods as discussed in Section 4.2. The final control signals are calculated via the difference between the control signals of each movement and its antagonistic movement for each DoF [44]. An example of the final control signals for DoF1 and DoF3 of subject six are illustrated in Figs. 10 and 11 using NMF and SNMF respectively.

### *5.3. Comparison through glove data reconstruction*

Synergies estimated by consTD, SNMF, and NMF were used to estimate the control signals from the testing EMG datasets. The glove data were reconstructed by applying ridge regression on the estimated testing control signals. This was done for each sensor of the 22 glove sensors where the ridge regression coefficients were calculated separately from the training data set as discussed in Section 4.3. An example of the four reconstructed glove data (sensor 12) plotted against the true glove data is shown in Fig. 13 for subject six.

For all subjects, $R^2$ were calculated between the true and reconstructed glove datasets for each wrist's DoF combination. The top three performing glove sensors were (8, 12, and 21) across all methods. The $R^2$ results for DoF1-3 are represented as a violin plot in Figs. 14 and 15 for datasets 1 and 2, respectively. The mean values for the three wrist's DoF combinations for both datasets are summarised in Table 1. The statistical analysis of two-sample *t*-test between the three methods (consTD, NMF, and SNMF) against random synergies showed that the three methods rejected the null hypothesis ($p \leq 0.05$). Hence, there is a significant difference between the $R^2$ results for the three methods and the randomly generated synergies.

## 6. Discussion

EMG has been used for decades to control prostheses [15]. Recently, several synergy-based systems have been proposed to achieve simultaneous and proportional myoelectric control [20,21]. These approaches rely on matrix factorisation methods to extract muscle synergies which are utilised to provide continuous control signals. However, those approaches are still limited in terms of the number of DoFs, task-dimensionality, and reliability. Because of the limitations of matrix factorisation methods, tensor decomposition was introduced to EMG signals for muscle synergies (i.e., modules) investigation [13,14,29]. Matrix factorisation might be suitable to extract spatial and temporal modules, but it cannot investigate the task-specific synergies. Hence,

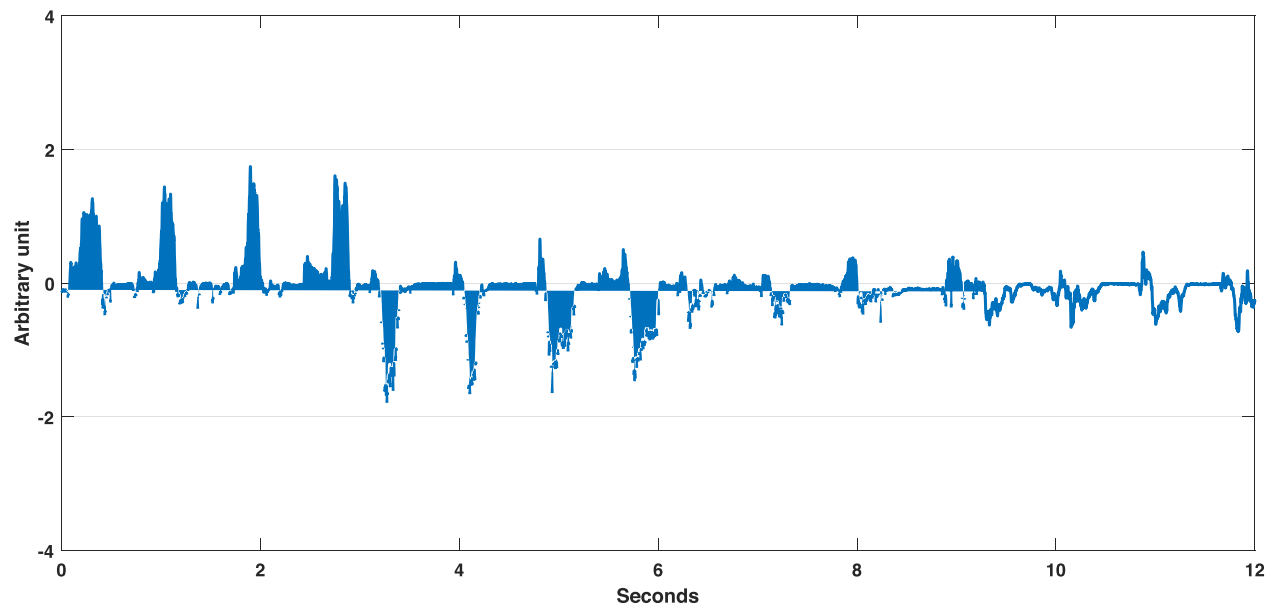

(a) *Final control signal for DoF1.*

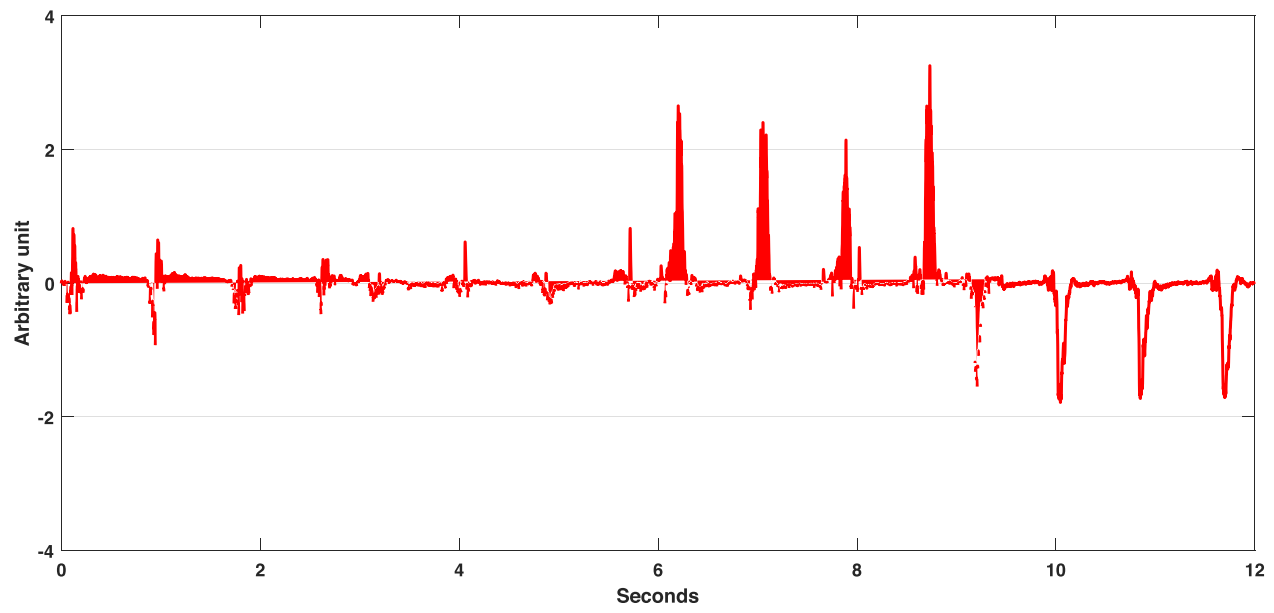

(b) *Final control signal for DoF3.*

**Fig. 12.** Final control signal for DoF1 (Panel a) and DoF3 (b) projected through direct projection of muscle synergies extracted via consTD recorded from subject six.

tensor decomposition could be suitable for muscle synergy applications in prosthesis control.

In this study, the potential application of higher-order tensor model in myoelectric control system was explored. We provided a scheme for applying synergies extracted via higher-order tensor decomposition in prosthesis control systems. This was approached by using a consTD method for synergy extraction from 3rd-order EMG tensor and incorporating the shared synergy concept. In an earlier study [14], we showed that the consTD method can estimate consistent synergies when the task dimensionality is increased up to 3-DoFs, while the traditional NMF was not able to extract consistent synergies when EMG segments were expanded to include additional DoFss. In addition, the consTD approach was better than NMF when the EMG data consists of several DoFs, since consTD includes shared synergies in the estimation process naturally [13]. Moreover, Takiyama et al. [29] showed that tensor decomposition enables the quantification of task-specific synergies in both spatial and temporal synergies simultaneously. While matrix factorisation methods including NMF can only quantify task-specific synergies in either spatial and temporal synergies when combined with a *posteriori* analysis [45,46]. Hence, we demonstrate here the ability of tensor decomposition in general and consTD specifically to estimate muscle synergies and control signals that can be utilised to provide a framework for simultaneous and proportional myoelectric control systems, especially with the increase of task-dimensionality and the number of DoFs.

A consTD scheme was proposed to estimate muscle synergies from training data for proportional myoelectric control. Muscle synergies were extracted via both NMF and SNMF for comparison. The estimated synergies were used to obtain control signals for each DoF through direct projection of the EMG testing data. The three methods were able to estimate the control signals for each DoF that can be used in synergy-based myoelectric control systems. However, consTD was able to use all data in one 3rd-order tensor, unlike matrix factorisation models where the data is segmented for each DoF as shown in Fig. 2. Moreover, the consTD method provides more information by including additional shared synergies as shown in Fig. 7, where spatial synergies 5 and 6 are shared between the tasks of DoFs 1 and 3 respectively. In comparison, matrix factorisation (Figs. 8 and 9) methods can only provide synergies for each task separately without any regard to the underlying shared synergistic information between tasks and/or DoFs.

In addition, the concept of tensor decomposition can include more information by expanding the tensors to add additional related data to EMG signals. These additional information can enhance the performance of decomposition and synergy extraction. For example, tensor decomposition was applied to joint angle and EMG data to investigate task-specific synergies [29] in addition to spatial and temporal synergies. This was done simultaneously using tensor decomposition unlike matrix factorisation approaches where it is limited to only two variables. Other prosthesis control studies [47,48] used inertial measurement units (IMUs) along with EMG to improve classification accuracy and reduce the number of used electrodes, which is essential for the practicality of prosthesis control. Hence, it could be useful to incorporate both EMG and IMU data in a data fusion scheme based on tensor factorisation to extract the underlying information between them. This could further

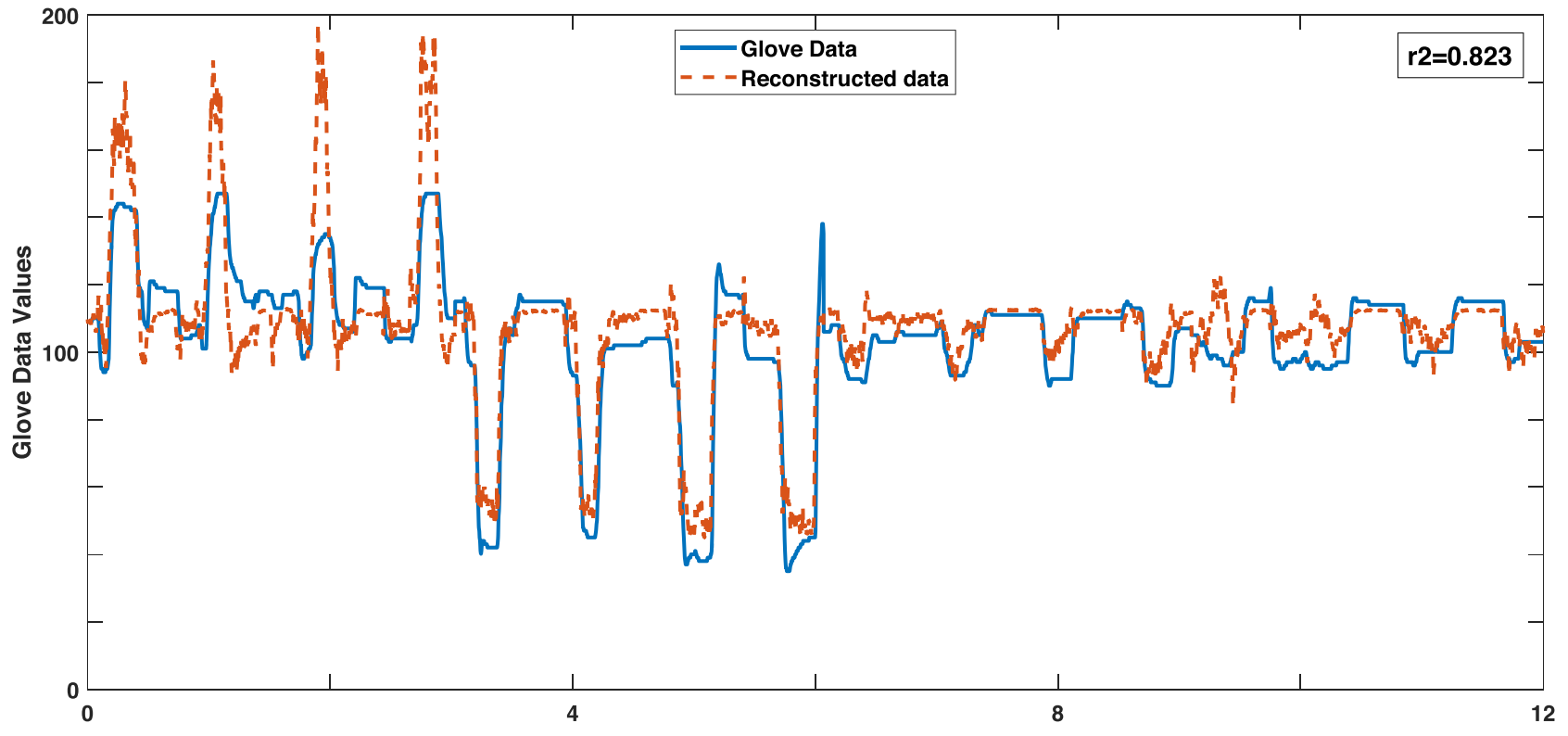

(a) *consTD.*

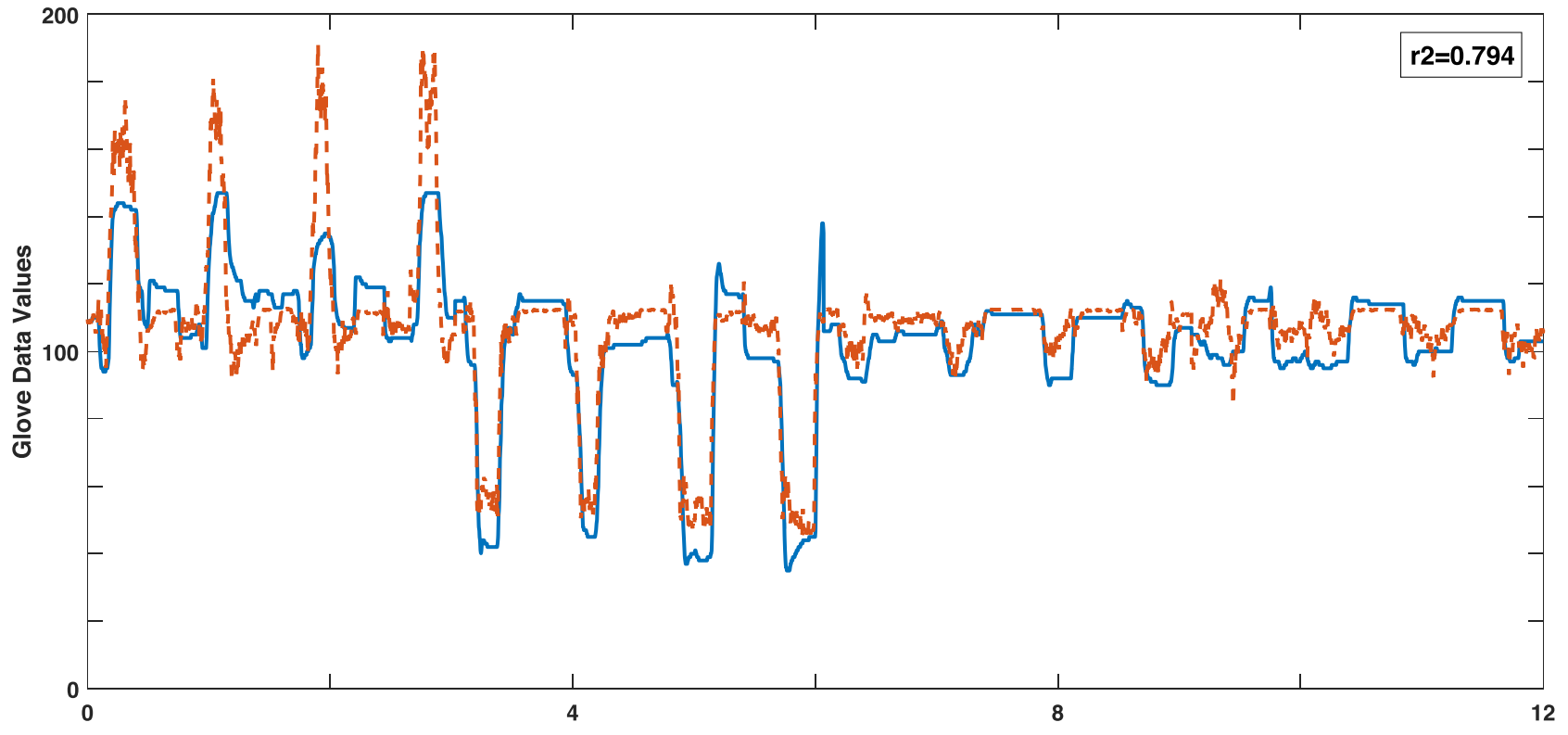

(b) *SNMF.*

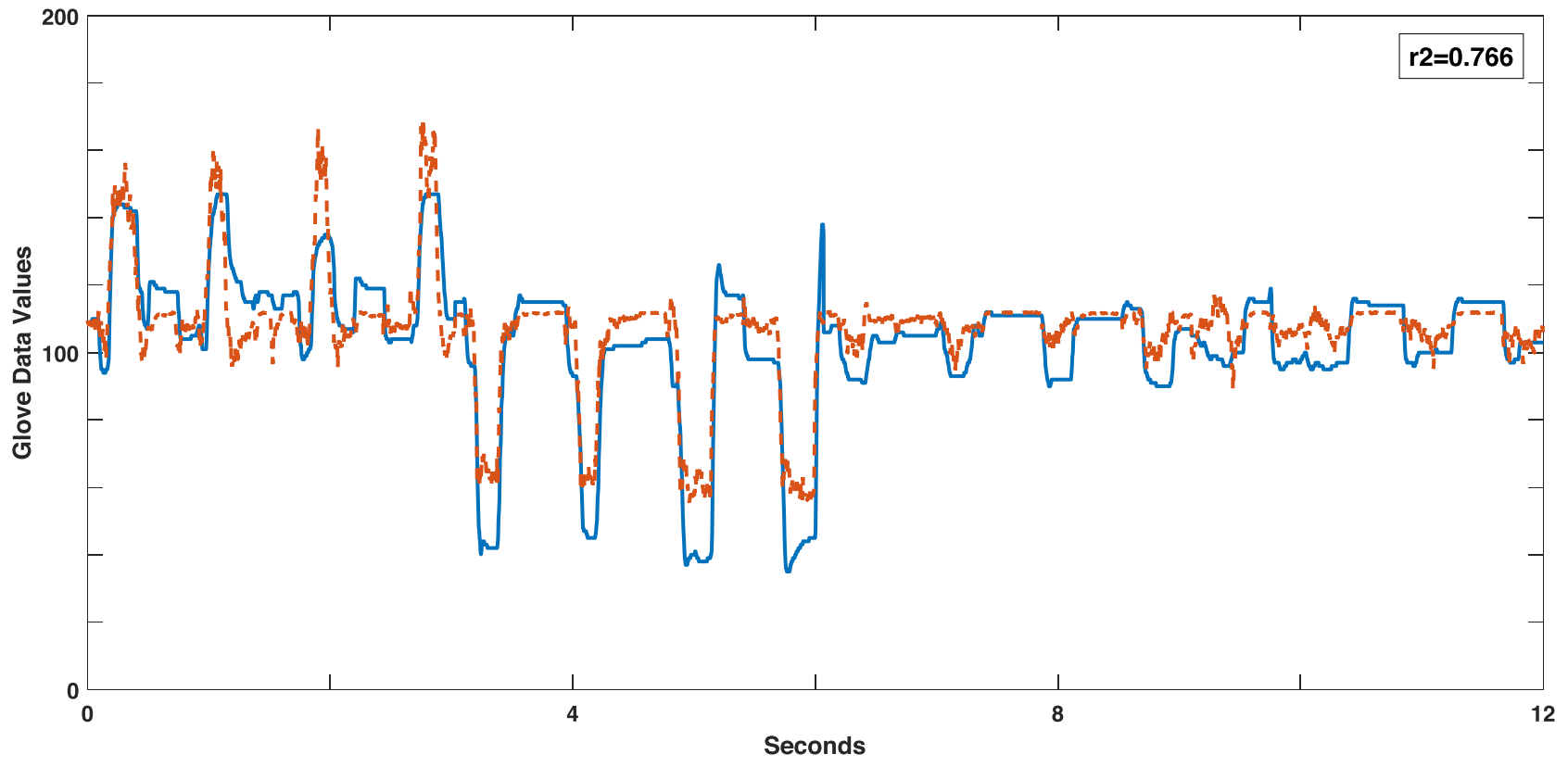

(c) *NMF.*

**Fig. 13.** Representative traces of wrist movement (DoF1-3) glove data (sensor 12) reconstruction using muscle synergies extracted via (a) consTD, (b) SNMF and (c) NMF synergies for subject 6.

improve the analysis and performance which cannot be done by the conventional methods.

The identified synergies and the extracted control signals were used to reconstruct the glove dataset through direct regression of the EMG testing data. The reconstructed glove data were compared with real glove data as shown in Fig. 13. $R^2$ was calculated between the reconstructed and actual glove data as a metric to assess each method. To rule out any statistical chance, random synergies were used to reconstruct glove data as well and two-sample *t*-test was performed on the three methods (NMF,SNMF and consTD). The statistical analysis of $R^2$ showed that there is a significant difference between the $R^2$ results of the three methods and the randomly generated synergies.

The reconstructed glove data that was computed via consTD method has higher $R^2$ values than that of the matrix factorisation methods as shown in Figs. 15 and 14. However, the $R^2$ values difference is not statistically significant. This is because ridge regression affected $R^2$ values. As a result, the differences between methods are not represented effectively. Another drawback is that the average $R^2$ value across all subjects for the three methods was generally modest. This is due to the fact that glove data may be not the best way to capture the hand kinematics, especially the wrist's DoF, as they rely on resistive bend-sensing [32].

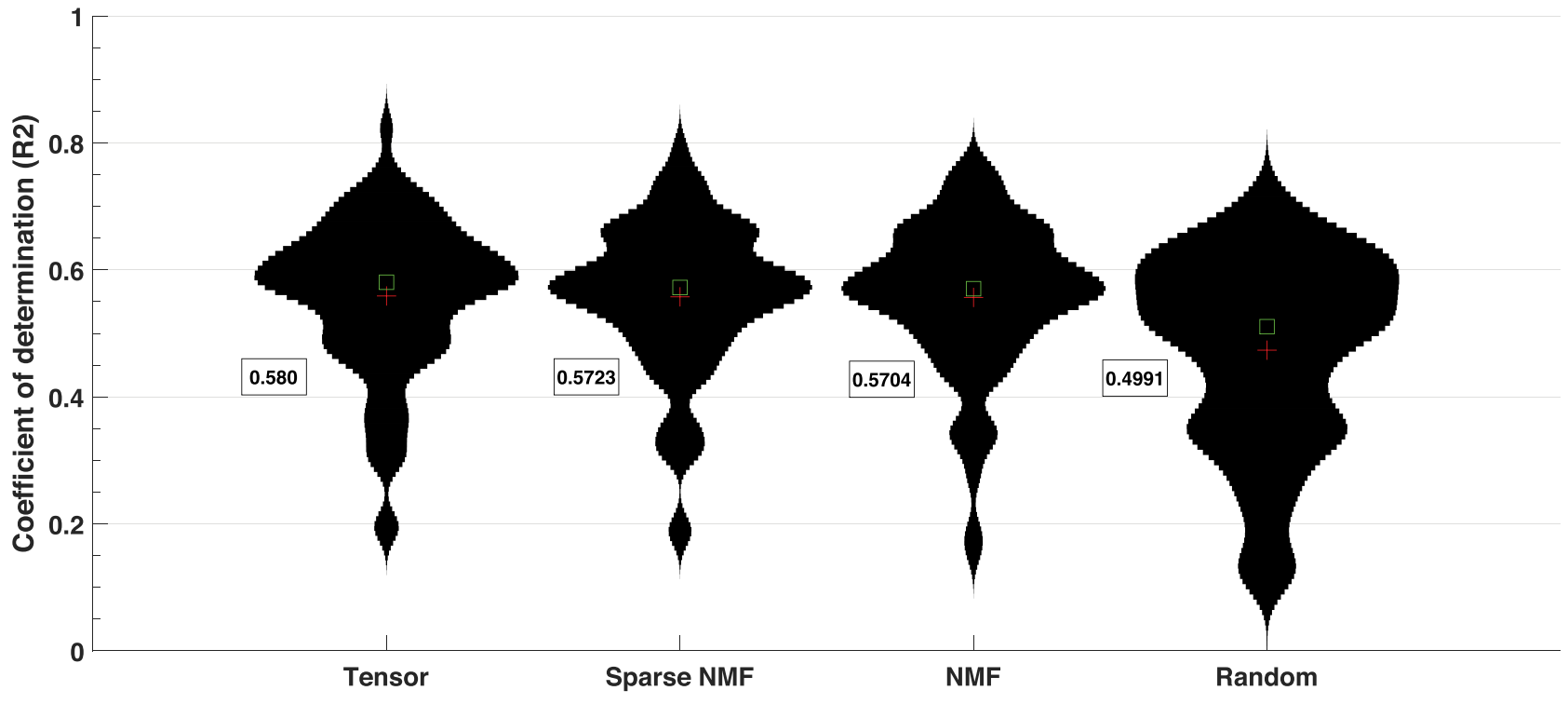

**Fig. 14.** Violin graph for the $R^2$ values of reconstructed glove data (DoF1-3) for each method across all subjects and top 3 sensors (8, 12 and 21). The mean and median are represented in the figure as red crosses and green squares respectively for dataset (1).

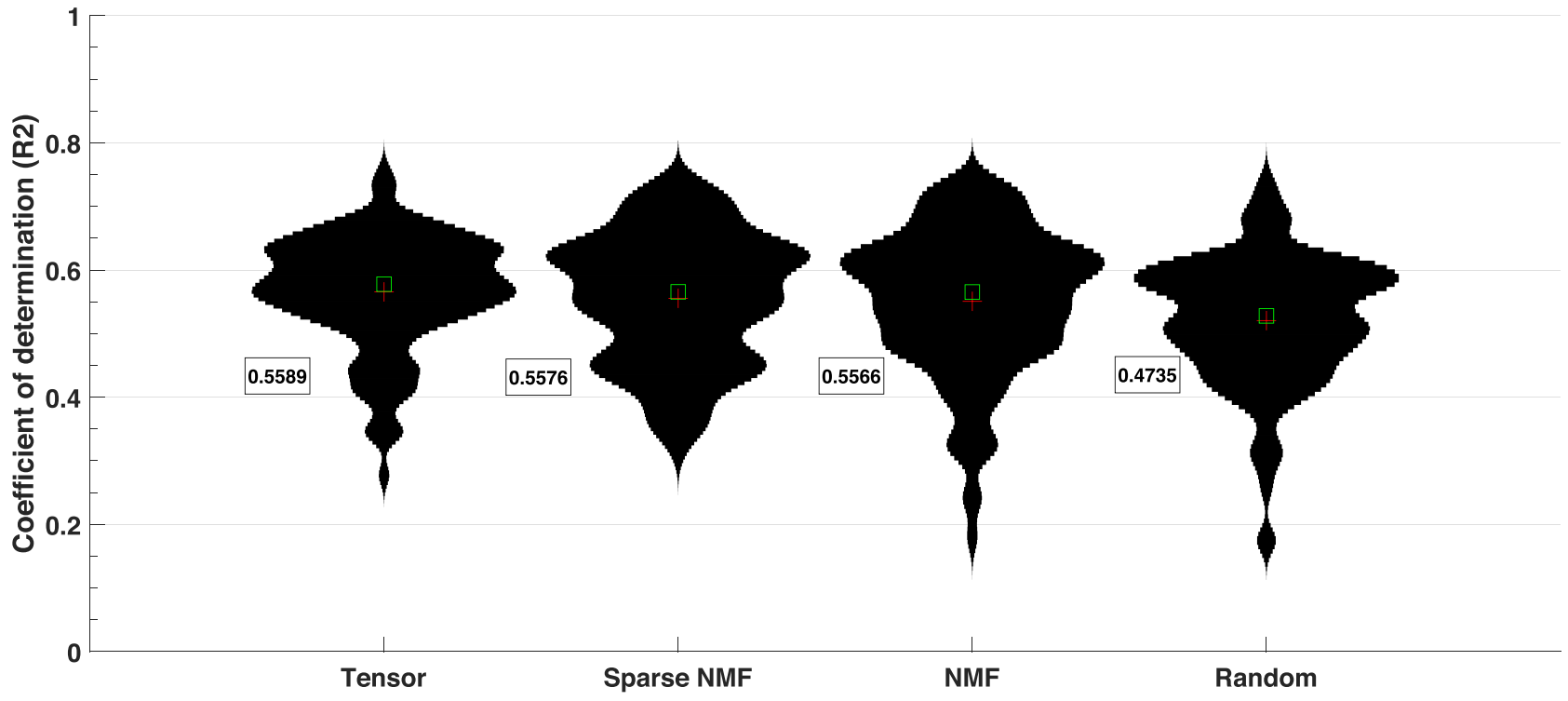

**Fig. 15.** Violin graph for the $R^2$ values of reconstructed glove data (DoF1-3) for each method across all subjects and top 3 sensors (1, 11 and 22). The mean and median are represented in the figure as red crosses and green squares respectively for dataset (2).

**Table 1**
The mean values of $R^2$ for the reconstructed glove data of the 3 DoFs combination.

| | | consTD | SNMF | NMF |
|---|---|---|---|---|
| DoF 1-2 | dataset-1 | 0.5241 | 0.5238 | 0.5146 |
| | dataset-2 | 0.5112 | 0.5111 | 0.4964 |
| DoF 1-3 | dataset-1 | 0.580 | 0.5723 | 0.5704 |
| | dataset-2 | 0.5589 | 0.5576 | 0.5566 |
| DoF 2-3 | dataset-1 | 0.535 | 0.541 | 0.532 |
| | dataset-2 | 0.516 | 0.512 | 0.511 |

However, this study provides a proof of concept for the use of higher-order tensor decomposition in proportional myoelectric control. For this application, 3rd-order tensor provides an easier approach to identify synergies for each DoF by adding this information to the tensor construction and decomposition. On the other hand, NMF methods have to extract synergies separately through DoF-wise training [20,24]. SNMF extracted synergies from two DoFs datasets [21], but another step was needed to identify synergies for each DoF after the factorisation process.

## 7. Conclusion

In summary, the novel consTD was presented as a method for synergy-based proportional myoelectric control. Tensor decomposition has not been utilised in any myoelectric control system. consTD was compared with NMF and SNMF methods, the current benchmarks in synergy-based myoelectric control schemes. The wrist's three main DoFs from two publicly available datasets were investigated in this comparison. Synergies extracted by the three methods (consTD, NMF, and SNMF) were used to estimate the control signal for each DoF through direct projection, to provide a proof of concept for the application of consTD in proportional myoelectric control. Then, the control signals were used to reconstruct the glove testing dataset for comparison. Although the consTD method is not significantly better than matrix factorisation techniques, its $R^2$ tends to be higher and it allows avoiding some of the problems associated with the training of the alternatives based on matrix factorisation. Therefore, we expect consTD to be a method worthy of further investigation to obtain myoelectric control signal.

## Authors' contribution

Ahmed Ebied: conceptualisation, methodology, software, investigation, writing; Eli Kinney-Lang: conceptualisation, methodology; Javier Escudero: conceptualisation, methodology, supervision.

## Conflict of interest

None declared.

## References

[1] A. d'Avella, M. Giese, Y.P. Ivanenko, T. Schack, T. Flash, Editorial: Modularity in motor control: from muscle synergies to cognitive action representation, Front. Comput. Neurosci. 9 (2015) 126.

[2] A. d'Avella, P. Saltiel, E. Bizzi, Combinations of muscle synergies in the construction of a natural motor behavior, Nat. Neurosci. 6 (2003) 300–308.
[3] M. Coscia, P. Tropea, V. Monaco, S. Micera, Muscle synergies approach and perspective on application to robot-assisted rehabilitation, in: R. Colombo, V. Sanguineti (Eds.), Rehabilitation Robotics, Elsevier, 2018, pp. 319–331 (Chapter 23).
[4] M.C. Tresch, P. Saltiel, E. Bizzi, The construction of movement by the spinal cord, Nat. Neurosci. 2 (1999) 162–167.
[5] D. Torricelli, F. Barroso, M. Coscia, C. Alessandro, F. Lunardini, E. Bravo Esteban, A. D'Avella, Muscle synergies in clinical practice: theoretical and practical implications, in: J.L. Pons, R. Raya, J. González (Eds.), Emerging Therapies in Neurorehabilitation II, vol. 10 of Biosystems & Biorobotics, Springer International Publishing, Cham, 2016, pp. 251–272.
[6] M.M. Nazifi, H.U. Yoon, K. Beschorner, P. Hur, Shared and task-specific muscle synergies during normal walking and slipping, Front. Human Neurosci. 11 (2017) 1–14.
[7] G. Martino, Y.P. Ivanenko, A. D'Avella, M. Serrao, A. Ranavolo, F. Draicchio, G. Cappellini, C. Casali, F. Lacquaniti, Neuromuscular adjustments of gait associated with unstable conditions, J. Neurophysiol. 114 (2015) p. jn.00029.2015.
[8] P. Saltiel, K. Wyler-Duda, A. D'Avella, M.C. Tresch, E. Bizzi, Muscle synergies encoded within the spinal cord: evidence from focal intraspinal NMDA iontophoresis in the frog, J. Neurophysiol. 85 (2001) 605–619.
[9] D.D. Lee, H.S. Seung, Learning the parts of objects by non-negative matrix factorization, Nature 401 (1999) 788–791.
[10] M.C. Tresch, V.C.-K.K. Cheung, A. D'Avella, Matrix factorization algorithms for the identification of muscle synergies: evaluation on simulated and experimental data sets, J. Neurophysiol. 95 (2006) 2199–2212.
[11] A. Ebied, E. Kinney-Lang, L. Spyrou, J. Escudero, Evaluation of matrix factorisation approaches for muscle synergy extraction, Med. Eng. Phys. 57 (2018) 51–60.
[12] A. Ebied, L. Spyrou, E. Kinney-Lang, J. Escudero, On the use of higher-order tensors to model muscle synergies, in: 2017 39th Annual International Conference of the IEEE Engineering in Medicine and Biology Society (EMBC), IEEE, 2017, pp. 1792–1795.
[13] A. Ebied, E. Kinney-lang, L. Spyrou, J. Escudero, Muscle activity analysis using higher-order tensor models: application to muscle synergy identification, IEEE Access 7 (2018) 27257–27271.
[14] A. Ebied, E. Kinney-Lang, J. Escudero, Consistency of muscle synergies extracted via higher-order tensor decomposition towards myoelectric control, in: 2019 9th International IEEE/EMBS Conference on Neural Engineering (NER), IEEE, 2019, pp. 315–318.
[15] E. Biddiss, D. Beaton, T. Chau, Consumer design priorities for upper limb prosthetics, Disability Rehabil. Assist. Technol. 2 (2007) 346–357.
[16] P. Geethanjali, Myoelectric control of prosthetic hands: state-of-the-art review, Med. Dev. (Auckland, N.Z.) 9 (2016) 247–255.
[17] L.J. Hargrove, E.J. Scheme, K.B. Englehart, B.S. Hudgins, Multiple binary classifications via linear discriminant analysis for improved controllability of a powered prosthesis, IEEE Trans. Neural Syst. Rehabil. Eng. 18 (2010) 49–57.
[18] D. Farina, N. Jiang, H. Rehbaum, A. Holobar, B. Graimann, H. Dietl, O.C. Aszmann, The extraction of neural information from the surface EMG for the control of upper-limb prostheses: emerging avenues and challenges, IEEE Trans. Neural Syst. Rehabil. Eng. 22 (4) (2014) 797–809.
[19] N. Jiang, S. Dosen, K.-r. Muller, D. Farina, Myoelectric control of artificial limbs: is there a need to change focus? [In the spotlight], IEEE Signal Process. Mag. 29 (5) (2012) 150–152.
[20] N. Jiang, H. Rehbaum, I. Vujaklija, B. Graimann, D. Farina, Intuitive, online, simultaneous, and proportional myoelectric control over two degrees-of-freedom in upper limb amputees, IEEE Trans. Neural Syst. Rehabil. Eng. 22 (3) (2014) 501–510.
[21] C. Lin, B. Wang, N. Jiang, D. Farina, Robust extraction of basis functions for simultaneous and proportional myoelectric control via sparse non-negative matrix factorization, J. Neural Eng. 15 (2018) 026017.
[22] N. Jiang, K.B. Englehart, P.a. Parker, Extracting simultaneous and proportional neural control information for multiple-dof prostheses from the surface electromyographic signal, IEEE Trans. Biomed. Eng. 56 (2009) 1070–1080.
[23] C. Choi, J. Kim, Synergy matrices to estimate fluid wrist movements by surface electromyography, Med. Eng. Phys. 33 (2011) 916–923.
[24] J. Ma, N.V. Thakor, F. Matsuno, Hand and wrist movement control of myoelectric prosthesis based on synergy, IEEE Trans. Human-Mach. Syst. 45 (2015) 74–83.
[25] R. Prevete, F. Donnarumma, A. D'Avella, G. Pezzulo, Evidence for sparse synergies in grasping actions, Sci. Rep. 8 (2018) 616.
[26] D.D. Lee, H.S. Seung, Algorithms for non-negative matrix factorization. Advances in Neural Information Processing Systems, 2001, pp. 556–562.
[27] A. de Rugy, G.E. Loeb, T.J. Carroll, Are muscle synergies useful for neural control? Front. Comput. Neurosci. 7 (2013) 19.
[28] A. Cichocki, D. Mandic, A.H. Phan, C. Caiafa, G. Zhou, Q. Zhao, L. De Lathauwer, Tensor decompositions for signal processing applications from two-way to multiway component analysis, IEEE Signal Process. Mag. 32 (2014) 1–23.
[29] K. Takiyama, H. Yokoyama, N. Kaneko, K. Nakazawa, Speed-dependent and mode-dependent modulations of spatiotemporal modules in human locomotion extracted via tensor decomposition, Sci. Rep. 10 (2020) 1–15.
[30] M. Atzori, A. Gijsberts, C. Castellini, B. Caputo, A.-G.M. Hager, S. Elsig, G. Giatsidis, F. Bassetto, H. M“uller, Electromyography data for non-invasive naturally-controlled robotic hand prostheses, Sci. Data 1 (2014) 140053.
[31] M. Atzori, A. Gijsberts, I. Kuzborskij, S. Elsig, A.-G. Mittaz Hager, O. Deriaz, C. Castellini, H. Muller, B. Caputo, Characterization of a benchmark database for myoelectric movement classification, IEEE Trans. Neural Syst. Rehabil. Eng. 23 (2015) 73–83.
[32] M. Atzori, A. Gijsberts, S. Heynen, A.-G.M. Hager, O. Deriaz, P. van der Smagt, C. Castellini, B. Caputo, H. Muller, Building the Ninapro database: a resource for the biorobotics community, Proceedings of the IEEE RAS and EMBS International Conference on Biomedical Robotics and Biomechatronics (2012) 1258–1265.
[33] A. Gijsberts, M. Atzori, C. Castellini, H. M”uller, B. Caputo, Movement error rate for evaluation of machine learning methods for sEMG-based hand movement classification, IEEE Trans. Neural Syst. Rehabil. Eng. 22 (2014) 735–744.
[34] L.R. Tucker, Some mathematical notes on three-mode factor analysis, Psychometrika 31 (1966) 279–311.
[35] P. Comon, Tensors: a brief introduction, IEEE Signal Process. Mag. 31 (2014) 44–53.
[36] T.G. Kolda, B.W. Bader, Tensor decompositions and applications, SIAM Rev. 51 (2008) 455–500.
[37] A. Smilde, R. Bro, P. Geladi, Multi-Way Analysis with Applications in the Chemical Sciences, John Wiley & Sons, Ltd, Chichester, UK, 2004.
[38] P. Comon, X. Luciani, A.L.F. de Almeida, Tensor decompositions, alternating least squares and other tales, J. Chemometrics 23 (2009) 393–405.
[39] R. Sands, F.W. Young, Component models for three-way data: an alternating least squares algorithm with optimal scaling features, Psychometrika 45 (1980) 39–67.
[40] K. Devarajan, Nonnegative matrix factorization: an analytical and interpretive tool in computational biology, PLoS Comput. Biol. 4 (2008) e1000029.
[41] H. Kim, H. Park, Sparse non-negative matrix factorizations via alternating non-negativity-constrained least squares for microarray data analysis, Bioinformatics 23 (2007) 1495–1502.
[42] C.A. Andersson, R. Bro, The N-way toolbox for MATLAB, Chemometrics Intell. Lab. Syst. 52 (2000) 1–4.
[43] A. Krasoulis, S. Vijayakumar, K. Nazarpour, Evaluation of regression methods for the continuous decoding of finger movement from surface EMG and accelerometry, in: 2015 7th International IEEE/EMBS Conference on Neural Engineering (NER), IEEE, 2015, pp. 631–634.
[44] N. Jiang, T. Lorrain, D. Farina, A state-based, proportional myoelectric control method: online validation and comparison with the clinical state-of-the-art, J. Neuroeng. Rehabil. 11 (2014) 110.
[45] G. Torres-Oviedo, L.H. Ting, Subject-specific muscle synergies in human balance control are consistent across different biomechanical contexts, J. Neurophysiol. 103 (2010) 3084–3098.
[46] H. Yokoyama, T. Ogawa, N. Kawashima, M. Shinya, K. Nakazawa, Distinct sets of locomotor modules control the speed and modes of human locomotion, Sci. Rep. 6 (2016) 1–14.
[47] A. Krasoulis, I. Kyranou, M.S. Erden, K. Nazarpour, S. Vijayakumar, Improved prosthetic hand control with concurrent use of myoelectric and inertial measurements, J. Neuroeng. Rehabil. 14 (2017) 71.
[48] A. Krasoulis, S. Vijayakumar, K. Nazarpour, Multi-grip classification-based prosthesis control with two EMG-IMU sensors, IEEE Trans. Neural Syst. Rehabil. Eng. 28 (2020) 508–518.